\documentclass[%
aip,
 amsmath,amssymb,
 reprint,%
]{revtex4-1}

\usepackage{graphicx}
\usepackage{dcolumn}
\usepackage{bm}

\usepackage[utf8]{inputenc}
\usepackage[T1]{fontenc}
\usepackage{mathptmx}
\usepackage{etoolbox}

\makeatletter
\def\@email#1#2{%
 \endgroup
 \patchcmd{\titleblock@produce}{\frontmatter@RRAPformat}{\frontmatter@RRAPformat{\produce@RRAP{*#1\href{mailto:#2}
  {#2}}}\frontmatter@RRAPformat}{}{} }%
\makeatother

\renewrobustcmd{\bfseries}{\fontseries{b}\selectfont}
\renewrobustcmd{\boldmath}{}
\newrobustcmd{\B}{\bfseries}

\usepackage{xcolor}
\usepackage{hyperref}
\usepackage{amsmath} 
\hypersetup{colorlinks=true}
\usepackage{comment}
\usepackage{rotating} 
\usepackage{subcaption}
\DeclareMathOperator\erf{erf}
\DeclareMathOperator\erfc{erfc}

\begin{document}

\preprint{AIP/123-QED}

\title{Molecular dynamics of nondegenerate hydrogen plasma using improved Kelbg pseudopotential with electron finite-size correction}


%
\newcommand\JIHT{Joint Institute for High Temperatures, Izhorskaya 13 Bldg 2, Moscow 125412, Russia}
\newcommand\MIPT{Moscow Institute of Physics and Technology, Institutskiy Pereulok 9, Dolgoprudny, Moscow Region,
 141701, Russia}
\newcommand\IPCP{Federal Research Center of Problems of Chemical Physics and Medicinal Chemistry, Academician Semenov avenue 1, Chernogolovka, Moscow region, 142432, Russia}

\author{G.S. Demyanov}
\affiliation{\JIHT}
\affiliation{\MIPT}
\author{P.R. Levashov}
\affiliation{\JIHT}
\affiliation{\MIPT}

\date{\today}

\begin{abstract} 
This paper is devoted to semiclassical molecular dynamics simulation of nondegenerate hydrogen plasma using an improved Kelbg pseudopotential. The main novelty of our method is accounting for the finite size of electrons. This modification resolves the nonphysical cluster formation at temperatures below 50 kK, which was first reported by Filinov~\textit{et al.} [\href{https://doi.org/10.1103/PhysRevE.70.046411}{Phys. Rev. E \textbf{70}, 046411}]. However, the energy still appears to be underestimated at low temperatures, as indicated by comparisons with the recent path integral Monte Carlo calculations [\href{https://doi.org/10.1063/5.0219405}{Phys. Plasmas 31, 110501 (2024)}].
Using the presented method, we analyze the dependence of radial distribution functions, composition, ionization degree, energy, and pressure on the plasma coupling parameter, while maintaining a fixed degeneracy parameter. Additionally, we demonstrate the impact of incorporating long-range interactions on the energy $N$-dependence by utilizing the angular-averaged Ewald potential. Finally, we compute the thermodynamic limits for energy and pressure.
\end{abstract}

\maketitle

\section{Introduction}

Hydrogen, being the ``simplest'' substance, has been investigated in a vast number of studies (see review paper~\onlinecite{Bonitz:PP:2024} and references therein). Its thermodynamic properties can be parameterized by the coupling parameter, $\Gamma$, and degeneracy parameter, $\chi$, which are related to the Brueckner parameter, $r_s$ (see Eq.~\eqref{eq:GammaChiDef}). Notably, the hydrogen atom is one of the analytically solvable quantum systems. This capability enables the development of first-principles methods toward density functional theory \cite{Bonitz:PP:2024, PhysRevE.105.065204, PhysRevE.107.015206}. The main method for \textit{ab initio} simulation of hydrogen thermodynamic properties is path integral Monte Carlo (PIMC) \cite{Filinov_2001,PhysRevE.63.066404, Filinov_2019, 10.1063/5.0211407, https://doi.org/10.1002/ctpp.201100085, Pierleoni2006}. In principle, this approach allows achieving results with arbitrarily high accuracy. Some works therefore refer to such results as ``quasi-exact''~\cite{10.1063/5.0138955}. The most recent hydrogen equation of state obtained by PIMC is presented in Ref.~\onlinecite{Filinov:PhysRevE:2023}.

Despite the fairly broad applicability of PIMC methods, their main drawback is the high computational cost. On the one hand, this cost is due to the use of high-temperature factorization required for calculations at relatively low temperatures~\cite{Filinov:PhysRevE:2023}. As a result, the cost increases linearly with the number of factorizations.

On the other hand, studies often consider conditions under which the electronic subsystem is degenerate. Consequently, accounting for coordinate permutations in  partition function leads to a determinant form of density matrix~\cite{DORNHEIM20181}. This immediately raises the computational complexity of the method to $O(N^3)$. In addition, the determinant can be negative; that is known as the fermion sign problem~\cite{RevModPhys.94.015006}. Typically, this problem is addressed using the re-weighting method that substantially increases the statistical error of such calculations~\cite{Dornheim_2021}.

Nevertheless, in the case of very low degeneracy, the computational cost can be significantly reduced. At such conditions it is reasonable to use pseudopotentials that take into account quantum effects and perform semiclassical simulations~\cite{Filinov:PhysRevE:2004}. The most rigorous way for this purpose is based on the density matrix that satisfies the Bl\"{o}ch equation~\cite{Feynman:1972:SMS}. For example, one can obtain an electron–proton interaction pseudopotential (p/p) for a given temperature by summing over the electron levels of a hydrogen atom using the definition of the two-particle density matrix~\cite{MILITZER201688}.

In other words, one needs to compute the Slater sum~\cite{Butlitsky2008, Bonitz2004} to extract an \textit{ab initio} p/p. The calculation of the Slater sum in the first-order of perturbation theory leads to the well-known Kelbg~p/p~\cite{Kelbg:1963, Bonitz:ContrPlasPhys:2023}. Thus, this p/p was derived from first principles, which explains its usage in numerous studies~\cite{Ebeling:CPP:1999, Lavrinenko:2018, Filinov:PRE:2020, Filinov:CPP:2001}. 

Note that various non-rigorous modifications of the Coulomb potential are often used as pseudopotentials. For example, the attractive and repulsive Coulomb potentials in molecular dynamics (MD) can be replaced by a repulsive Debye potential~\cite{ANGEL20111812}. In Ref.~\onlinecite{Tiwari:PRE:2017}, a classical ultracold plasma is simulated with a ``pseudo-Coulomb'' p/p that contains a repulsive core at short distances with a parameter controlling the short-range behavior of the p/p. Moreover, this p/p is zero at zero distance. As shown in Ref.~\onlinecite{Tiwari:PRE:2017}, it leads to a strong dependence of the simulation results on the chosen p/p parameter. Nevertheless, the work presents simulations for both nearly ideal and strongly coupled plasmas, $\Gamma\leq 20$.

In Ref.~\onlinecite{PhysRevE.98.033307}, the behavior of a weakly coupled ($\Gamma\leq 0.14$) plasma is also investigated by MD with regularization of the Coulomb potential at short distances. In contrast to the previous study~\cite{Tiwari:PRE:2017}, the p/p has a finite value at zero distance that is equal to the ionization energy of an isolated hydrogen atom (13.6~eV). At distances smaller than a certain cutoff, the potential approaches this constant value quadratically as the distance tends to zero. It is shown that at low temperatures the degree of ionization is zero, whereas molecular formation is not reported. This feature may be a serious limitation of the chosen p/p. A strong sensitivity of the results to the p/p depth is also demonstrated.

In the recent study~\cite{PhysRevE.111.015204}, a more sophisticated interaction model between particles is employed. In this approach, electrons do not participate in MD simulation and are instead incorporated into effective interaction potentials between ions and neutral particles. The work considers only the $62.5$ kK isotherm to avoid molecular formation. Under these conditions, the degeneracy of the system varies from weak to moderate. Although such a model reproduces the characteristic minimum in the $r_s$-dependence of the ionization degree, it yields incorrect results at high densities. Furthermore, as previously mentioned, the model does not consider molecular formation that limits its applicability to temperatures above 52 kK.

Note that there exists a vast number of methods and studies devoted to the simulation of hydrogen properties. In addition to the PIMC and MD approaches mentioned above, there are quantum molecular dynamics based on density functional theory~\cite{Holst:PRB:2008,PhysRevB.101.195129, PhysRevE.93.063207, PhysRevLett.102.075002}, wave‑packet molecular dynamics~\cite{Lavrinenko:2016, Lavrinenko:2018, Lavrinenko:PRE:2021}, as well as various theoretical works on the virial expansion~\cite{PhysRevE.104.045204, Trigger_2023, Ebeling2025} and the hypernetted chain approximation \cite{Ichimaru:PhysRevA:1985, Ichimaru:PhysRep:1987}. 

The present study is devoted to first-principles simulations of a non-degenerate hydrogen plasma using MD within the framework of density matrix theory. To achieve this purpose, we employ the improved Kelbg p/p developed and presented by Filinov \textit{et al.} ~\cite{Filinov:PhysRevE:2004} This p/p accounts for low‑temperature effects in the Kelbg p/p that are absent in the original perturbative solution by G. Kelbg~\cite{Kelbg:1963}.

Nevertheless, Ref.~\onlinecite{Filinov:PhysRevE:2004} demonstrated that in simulations with the improved Kelbg p/p excessive particle attraction is observed at temperatures below 50~kK due to the appearance of bound states between electrons with identical spin projection. As a consequence, the total energy is significantly lower than the correct value.

The key distinction of the present paper, in comparison with the methodology of Ref.~\onlinecite{Filinov:PhysRevE:2004}, is an approximate account of the finite electron size that resolves the formation of nonphysical clusters (see Eq.~\eqref{eq:fdrgdfsef}). This enables the methods developed in Ref.~\onlinecite{Filinov:PhysRevE:2004} to be applied at temperatures lower than 50~kK. In this way, we compute the energy and pressure of a nondegenerate ($\chi = 10^{-2}$) hydrogen plasma, the composition and ionization degree, as well as the radial distribution functions, in the range $0.1 \leq \Gamma \leq 3$ of the coupling parameter $\Gamma$. Particular attention is paid to the impact of accounting for the long-range Coulomb interaction via the angular-averaged Ewald potential~\cite{Demyanov:JPA:2022, Demyanov:CPC:2024, Demyanov:PhysRevE:2022} on the $N$-convergence of the energy to the thermodynamic limit.

The article is organized as follows. Section~\ref{sec:hydrThem} represents interaction pseudopotentials, pair contributions to energy and pressure, and forces together with a modification for the electrons with the same spin projection. Section~\ref{sec:sim} discusses MD simulation parameters and the calculation of composition and ionization degree. We discuss our computational results in Section~\ref{sec:results}. Finally, we summarize our study in Section~\ref{sec:concl}.

\section{Hydrogen thermodynamics and interaction pseudopotentials \label{sec:hydrThem}}
We consider a system of $N_e$ electrons and $N_p = N_e$ protons. The masses of these particles are denoted as $m_e$ and $m_p$, respectively. The operator $\hat{H}$ represents the Hamiltonian of the system.  The particles are placed in a cubic cell of volume $L^3$ at coordinates $\textbf{r}_i$, where $i = 1,\ldots, N$.  Periodic boundary conditions are applied, resulting in each $i$th particle having an infinity number of images located at coordinates $\textbf{r}_i~+~\bm{\eta}L$, where $\bm{\eta} \in \mathbb{Z}^3$ is the integer vector. The symbol $\textbf{R} = (\textbf{r}_1, \ldots, \textbf{r}_N)$ denotes all particle positions.

Thermodynamic properties at an inverse temperature $\beta~=~(k_BT)^{-1}$ (where $k_B$ is the Boltzmann constant) can be specified by the coupling $\Gamma$ and degeneracy parameter~$\chi$:
\begin{equation}
	\label{eq:GammaChiDef}
	\Gamma = e^2\beta/r_a, \quad \chi = n_e\Lambda^3,
\end{equation}
where $r_a = (4\pi N_e/3)^{-1/3}L$ is the mean interparticle distance between the electrons; $n_e = N_e / L^3$ is the \emph{electron} density; \mbox{$\Lambda = (2\pi \hbar^2 \beta/m_e)^{1/2}$} is the electron thermal de Broglie wavelength. One may also use the pair $(r_s, \theta)$ as an alternative to $(\Gamma, \chi)$, where $r_s = r_a / a_B$ is the Brueckner parameter, and $\theta = (\beta E_f)^{-1}$ is the reduced temperature. Here, $E_f = \frac{\hbar^2}{2m_e}(3\pi^2N_e/L^3)^{2/3}$ denotes the Fermi energy and $a_B = \hbar^2 / (m_e e^2)$ is the Bohr radius. Also,  $E_H = m_e e^4/\hbar^2$ is the Hartree energy. 

We should also specify two characteristic energy scales for a hydrogen plasma. First, the ionization energy of the hydrogen atom is 13.6 eV, or in temperature units $T_H = 157.8$ kK. Second, the dissociation energy of the hydrogen molecule is 4.52 eV~\cite{Bonitz:PP:2024}, i.e., $T_{H_2} = 52.4$ kK.

In this paper, a plasma is called non-degenerate if $\chi \leq 0.01$. To perform an MD simulation of a non-degenerate hydrogen plasma, it is necessary to obtain interparticle pseudopotentials and  interaction forces~\cite{Demyanov:CPC:2024}, as well as contributions to the total energy and pressure. Next we consider an approach based on solving the Bl\"{o}ch equation for the density matrix, from which the necessary functions can be extracted.

\subsection{Density matrix and improved Kelbg pseudopotential}
The density matrix $\hat\rho(\beta)$ and its coordinate representation $\rho(\textbf{R}, {\textbf{R}'}; \beta)$ can be used to compute the thermodynamic properties of a quantum system at a temperature $T$~\cite{Feynman:1972:SMS}:
\begin{equation}
	\hat\rho(\beta) = \exp(-\beta \hat H),\quad  \rho(\textbf{R}, {\textbf{R}'}; \beta) = \langle \textbf{R} |\hat{\rho}(\beta)| {\textbf{R}'}\rangle.
	\label{eq:denmat}
\end{equation}
The density matrix satisfies the following Bl\"{o}ch equation:
\begin{equation}
	\label{eq:bloch}
	\frac{d \hat\rho(\beta)}{d \beta} = -\hat H\hat\rho(\beta), \quad \hat\rho(0) = \hat{1}.
\end{equation}
Note that the density matrix $\hat\rho(\beta)$ does not take into account the indistinguishability of particles. Then the statistical sum $Z(\beta)$ takes the following form:
\begin{equation}
	\label{eq:part_func1}
	Z(\beta) = \frac{1}{N!}\mathrm{Sp} \hat{\rho}(\beta) = \frac{1}{N!}\int d\textbf{R} \rho(\textbf{R}, {\textbf{R}}; \beta),
\end{equation}
where the factor $1/N!$ takes into account all possible permutations of particles.

If the Hamiltonian consists only of a kinetic term, i.e. there is no interaction between particles, one can obtain the following expression for any $\beta~>~0$~\cite{Feynman:1972:SMS}:
\begin{equation}
	\rho_0(\textbf{R}, {\textbf{R}'}; \beta)
	=
	\prod_{i = 1}^{N}
	\left[\left(\tfrac{m_i}{2\pi\hbar^2\beta}\right)^{3/2}\exp\left(-\tfrac{m_i(\textbf{r}_i-\textbf{r}'_i)^2}{2\hbar^2\beta}\right)\right].
\end{equation}
Let the potential energy operator have the following form:
\begin{equation}
	\hat{U} = \frac{1}{2}\sum_{i=1}^{N}\sum_{\substack{j = 1\\ j\neq i}}^{N}q_iq_j\phi(\textbf{r}_{ij}),
\end{equation}
where $q_i$ is the particle charge.
Next we expand the interaction potential $\phi(\textbf{r}_{ij})$ into a Fourier integral:
\begin{equation}
	\phi(\textbf{r}_{ij}) = \frac{1}{(2\pi)^3}\,\int d\textbf{k} \phi(\textbf{k})e^{i\textbf{k}\cdot \textbf{r}_{ij}}.
\end{equation}
Here, $\phi(\textbf{k})$ is the Fourier component of the potential $\phi(\textbf{r}_{ij})$. The solution of Bl\"{o}ch equation~\eqref{eq:bloch} in the first order of perturbation theory, i.e. in the limit $\beta\to 0$ or $\Gamma \to 0$, is written as follows (see details of the derivation in Refs.~\onlinecite{Kelbg:1963, Demyanov_Levashov_2022}):
\begin{equation}
	\label{eq:dmKelbg}
	\langle \textbf{R} |\hat{\rho}(\beta)| {\textbf{R}'}\rangle
	=
	\rho_0(\textbf{R}, {\textbf{R}'}; \beta)
	\exp\left[
	-
	\tfrac{\beta}{2}\sum_{\substack{i,j = 1\\ j\neq i}}^{N}q_iq_j \Phi(\textbf{r}_{ij},\textbf{r}'_{ij}; \beta)\right],
\end{equation}
where $\Phi(\textbf{r}_{ij},\textbf{r}'_{ij};\beta)$ denotes the Kelbg functional:
\begin{equation}
	\label{eq:kelbgfuncDef}
	\Phi(\textbf{r}_{ij},\textbf{r}'_{ij}; \beta) = {}\\
	\frac{1}{8\pi^3}
	\int\limits_0^{1}d\alpha
	\int {\phi}(\textbf{k})
	e^{
		i \textbf{k}\cdot\textbf{d}_{ij}(\alpha)  
	}e^{
	-\alpha(1-\alpha)
	\lambda_{ij}^2 k^2 
}
	d\textbf{k}
\end{equation}
with diagonal elements:
\begin{equation}
	\label{eq:diag_kelbg_gen}
	\Phi(\textbf{r}_{ij},\textbf{r}_{ij};\beta) = 
	\frac{1}{4\pi^3}
	\int  {\phi}(\textbf{k})
	e^{
		i \textbf{k}\cdot \textbf{r}_{ij}
	}\frac{D(\lambda_{ij} k/2)}{\lambda_{ij} k}d\textbf{k},
\end{equation}
where $\textbf{d}_{ij}(\alpha) = \alpha\textbf{r}_{ij}+(1-\alpha)\textbf{r}'_{ij}$ and $\textbf{r}'_{ij} = \textbf{r}'_{i}-\textbf{r}'_{j}$.
Here, $\lambda_{ij}=\lambda_{ij}(\beta) = \sqrt{\hbar^2\beta/(2\mu_{ij})} $ is the reduced thermal de Broglie  wavelength, $\mu_{ij}^{-1} = m_i^{-1} + m_j^{-1}$ denotes the reduced mass,  and $D(x)$ is the Dawson function.

In the case of Coulomb potential, $\phi(\textbf{k}) = 4\pi/k^2$, Kelbg's solution has the following form~\cite{Filinov_2001}:
\begin{equation}
\label{eq:kelbgpseudo}
\Phi \to \Phi_0(\textbf{r}_{ij},\textbf{r}'_{ij};\beta) =
\int\limits_0^{1}
\frac{d\alpha}{ d_{ij}(\alpha)} \mathrm{erf}\left(\frac{d_{ij}(\alpha) / \lambda_{ij}(\beta)}{2\sqrt{\alpha(1-\alpha)}}\right)
\end{equation}
with diagonal elements $\Phi_0(\textbf{r}_{ij},\textbf{r}_{ij};\beta)\equiv \Phi_0(r_{ij};\beta)$~\cite{Kelbg:1963}:
\begin{equation}
	\label{eq:diagkelbgtheta0}
	\Phi_0(\textbf{r}_{ij},\textbf{r}_{ij};\beta) \\=  \frac{1}{ r_{ij}}K_0(r_{ij}/\lambda_{ij}(\beta)),
\end{equation}
where 
\begin{equation}
	K_0(x) = 1 - e^{-x^2} + x\sqrt{\pi} \mathrm{erfc}(x).
\end{equation}

The Kelbg p/p is relevant solely when the difference between the potential energy derived from the function $\Phi(\textbf{r},\textbf{r}; \beta)$ and the classical potential energy obtained from the potential $\phi(\textbf{r})$ is small, due to the use of perturbation theory in solving the Bl\"{o}ch equation. Consequently, the Kelbg p/p inaccurately represents bound states, i.e., short-range interaction behavior at low temperatures. For example, it fails to account for the formation of hydrogen molecules. However, it can be directly utilized to simulate a system at $\Gamma \ll 1$, representing almost ideal gas~\cite{Demyanov:CPC:2024}.

To take into account temperature effects in the density matrix, it is necessary to obtain the p/p directly from the exact solution of the Bl\"{o}ch equation. However, it is convenient not to solve the Bl\"{o}ch equation but to use the following property of the density matrix (for two particles):
\begin{equation}
	\label{eq:gfnhgrsfgsef}
	\rho(\textbf{r}, \textbf{r}';\beta) = \int d \textbf{r}'' \rho(\textbf{r}, \textbf{r}'';\beta/2)\rho(\textbf{r}'', \textbf{r}';\beta/2).
\end{equation}
Thus, one can calculate the density matrix at a temperature $T$ using the density matrix at the temperature $2T$. In this way, the Kelbg solution (or a simpler expression) can serve as an initial approximation for the interaction part of the density matrix, allowing for the numerical computation of the exact density matrix for any $T~>~0$. 

This procedure was carried out by Filinov~\textit{et~al.} in Ref.~\onlinecite{Filinov:PhysRevE:2004}. In order to make the obtained dependencies convenient to use, a temperature parameter $\gamma_{ij}(\beta)$ was introduced into the original Kelbg p/p:
\begin{equation}
	\label{eq:gtdfesbtdgfr}
	\Phi^\text{I}_0(\textbf{r}_{ij},\textbf{r}_{ij};\beta) \equiv \Phi^\text{I}_0(r_{ij};\beta) =  \frac{1}{r_{ij}} K_0^\text{I}(r_{ij}/\lambda_{ij}(\beta), \gamma_{ij}(\beta))
	,
\end{equation}
where the improved function $K_0(x)$ is the following~\cite{Filinov:PhysRevE:2004}:
\begin{equation}
	K_0^\text{I}(x, \gamma_{ij}(\beta))=
	1 - e^{-x^2}+ \frac{x\sqrt{\pi}}{\gamma_{ij}(\beta)} \mathrm{erfc}\left(x\gamma_{ij}(\beta)\right),
\end{equation}
to approximate the obtained exact density matrix. Equation~\eqref{eq:gtdfesbtdgfr} defines the improved Kelbg p/p~\cite{Filinov:PhysRevE:2004}. The parameter $\gamma_{ij}(\beta)$ is a temperature function that depends on the types of interacting particles $i, j$ and $\gamma_{ij}(\beta \to 0) \to 1$. Next, $\Phi^\text{I}_{0,ee}$ and $\Phi^\text{I}_{0,ep}$ denote the improved Kelbg p/p for electron-electron and electron-proton interactions.
The explicit dependence of $\gamma_{ij}(\beta)$ on temperature for electron-proton and electron-electron interactions is given in equations (22) and (23), respectively, in the work~\onlinecite{Filinov:PhysRevE:2004}.

Thus, the improved Kelbg p/p accounts for low-temperature effects. However, to ensure the stability of a two-component system even at low degeneracy, it is necessary to consider the statistics of electrons, particularly their exchange interaction~\cite{Lieb:AM:1972, Baus:PR:1980}. Following Ref.~\onlinecite{Filinov:PhysRevE:2004}, we will obtain the exchange corrections to the electron-electron interaction.

\subsection{Electron exchange interaction}
As exchange effects are associated with the Pauli exclusion principle, it is necessary to account for the symmetry of a two-particle density matrix. Two electrons can be in a triplet state or in a singlet state, which corresponds to the same or opposite spin projections, respectively.  Then the diagonal elements of the two-electron density matrix in the singlet and triplet states are written as follows~\cite{Filinov:PhysRevE:2004, Feynman:1972:SMS}:
\begin{multline}
	\label{eq:fumygtdfr}
	\rho^\text{S(T)}_{ee}((\textbf{r}_1, \textbf{r}_2), (\textbf{r}_1, \textbf{r}_2); \beta)
	=
	\frac{1}{2!}\Bigg[\rho_{ee}((\textbf{r}_1, \textbf{r}_2), (\textbf{r}_1, \textbf{r}_2); \beta) \\ \pm  \rho_{ee}((\textbf{r}_1, \textbf{r}_2), (\textbf{r}_2, \textbf{r}_1); \beta)\Bigg] \equiv \tfrac{1}{2} \left(\tfrac{m_e}{2\pi\hbar^2\beta}\right)^{3}e^{-\beta e^2 \Phi^\text{S(T)}_{ee}(\textbf{r}_{12};\beta)},
\end{multline}
where the sign <<$+$>> corresponds to the singlet (S) state (the coordinate part of the two-electron wave function is symmetric), and the sign <<$-$>> corresponds to the triplet (T) state (the coordinate part of the two-electron wave function is antisymmetric). Here, $\rho_{ee}$ is the density matrix for two distinguishable electrons and is a solution of Eq.~\eqref{eq:bloch}. The last expression is a Boltzmann-type density matrix, which defines a new p/p \(\Phi^\text{S(T)}_{ee}(\textbf{r}_{12};\beta)\). This electron-electron p/p accounts for both the Coulomb interaction and Fermi statistics. 
Substituting the density matrix from Eq.~\eqref{eq:dmKelbg} with the improved Kelbg p/p~\eqref{eq:gtdfesbtdgfr} into Eq.~\eqref{eq:fumygtdfr}, we obtain (see also Eq.~(14) in Ref.~\onlinecite{Filinov:PhysRevE:2004}):  
\begin{multline}
	\label{eq:phi_st}
	\exp\left(-\beta e^2 \Phi^\text{S(T)}_{0, ee}(\textbf{r}_{12};\beta)\right)
	\\=
	e^{-\beta e^2 \Phi^\text{I}_{0, ee}(\textbf{r}_{12},\textbf{r}_{12}; \beta) }
	\pm \exp\left(-\frac{m_e r^2_{12}}{\hbar^2 \beta} \right)
	e^{-\beta e^2 \Phi^\text{I}_{0, ee}(\textbf{r}_{12},-\textbf{r}_{12}; \beta)}.
\end{multline}

As can be seen, the second term containing the non-diagonal p/p $\Phi^\text{I}_{0, ee}(\textbf{r}_{12},-\textbf{r}_{12}; \beta)$ can be associated with the coordinate exchange.
It is stated in Ref.~\onlinecite{Filinov:PhysRevE:2004} that the following approximation can be used:
\begin{equation}
	\label{eq:dhtgrfe}
	\Phi^\text{I}_{0, ee}(\textbf{r}_{12},-\textbf{r}_{12}; \beta) \approx \Phi^\text{I}_{0, ee}(\textbf{r}_{12},\textbf{r}_{12}; \beta),
\end{equation}
based upon the approximation of the non-diagonal p/p by diagonal components~\cite{Filinov_2001, Filinov:PhysRevE:2004}:
\begin{equation}
	\label{eq:dhyfhgdgtg}
	\Phi(\textbf{r},\textbf{r}'; \beta) \approx \frac{1}{2}\left(\Phi(\textbf{r},\textbf{r}; \beta)+\Phi(\textbf{r}',\textbf{r}'; \beta)\right).
\end{equation}
Note that in the case of a high-temperature solution, the approximation~\eqref{eq:dhyfhgdgtg} is quite accurate. However, in the case of low temperatures, Eq.~\eqref{eq:dhyfhgdgtg} may not be valid~\cite{Ebeling_2006}, especially at small distances.

Solving Eq.~\eqref{eq:phi_st} together with approximation~\eqref{eq:dhtgrfe}, one finds the necessary p/p that takes into account low-temperature effects and Fermi statistics:
\begin{equation}
	\label{eq:frddtgrf}
	\Phi^\text{S(T)}_{0, ee}({r};\beta) = \Phi^\text{I}_{0, ee}({r}; \beta) -\tfrac{1}{\beta e^2} \ln \left(1 \pm \exp\left[-\tfrac{m_e r^2}{\hbar^2 \beta}\right]\right),
\end{equation}
which is the same as Eq.~(17) in Ref.~\onlinecite{Filinov:PhysRevE:2004}.
Thus, at small distances, $|\textbf{r}|\to 0$, the additional contribution in the case of a singlet state tends to a constant ($\ln2$), while it diverges as $\ln r$ in the case of a triplet state.

\begin{figure}
	\centering
	\includegraphics[width=1\linewidth]{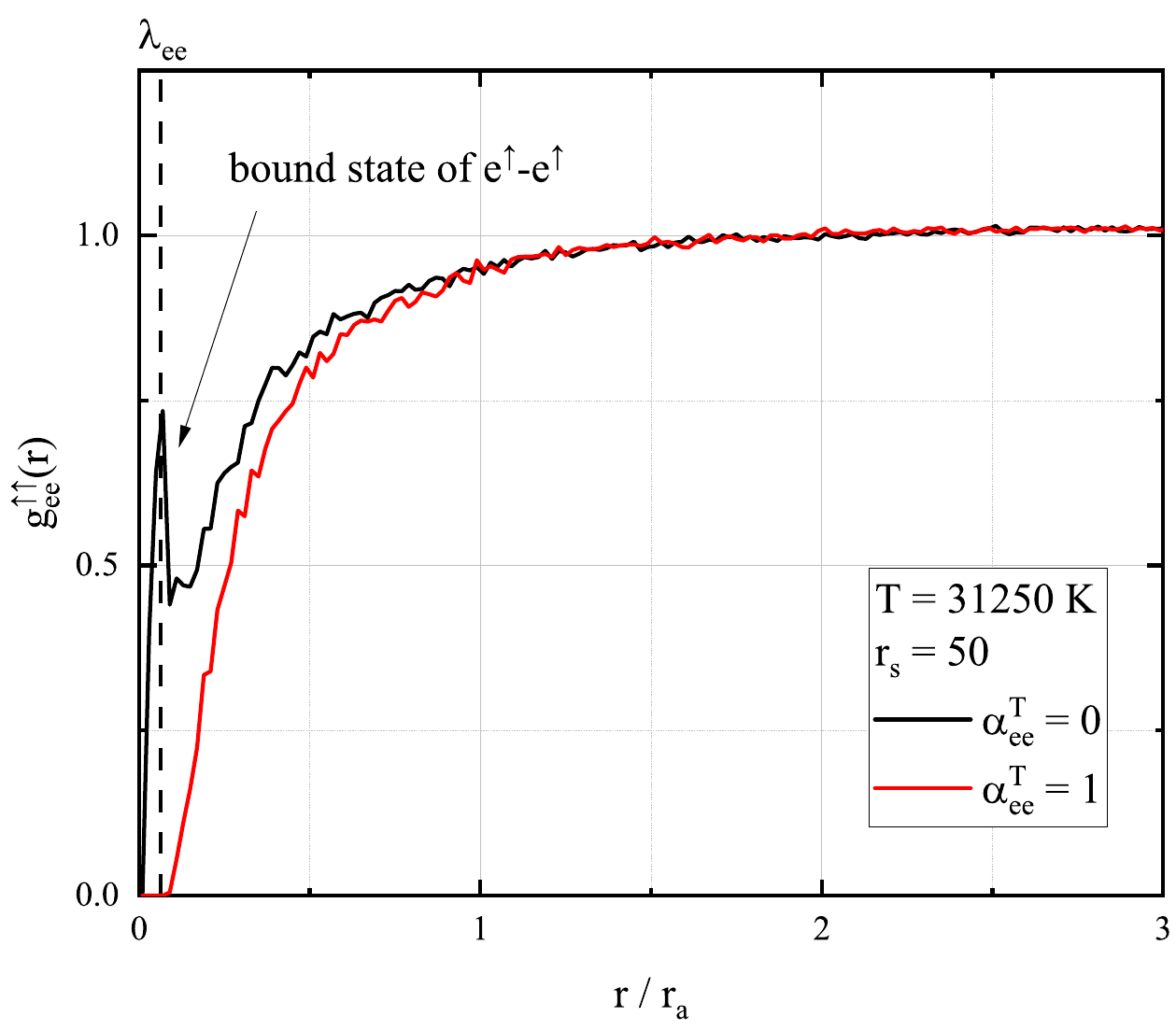}
	\caption{An example of electrons with identical spin projections bound states formation. The black curve is obtained without modifying the forces ($\alpha^\text{T}_{ee}=0$), and the red curve was obtained using~\eqref{eq:fdrgdfsef} with the parameter $\alpha^\text{T}_{ee}=1$.}
	\label{fig:spineebad}
\end{figure}

However, as simulations in Refs.~\onlinecite{Filinov:PhysRevE:2004, Golubnychiy2004} show, logarithmic repulsion at $r\ll \lambda_{e}$ in Eq.~\eqref{eq:frddtgrf} is insufficient to prevent electrons with the \textit{same} spin projection from forming a bound state in a ``molecule.'' This phenomenon occurs because the presence of a proton in an atom weakens the repulsion between electrons that have the same spin projection. As a result, these electrons can approach each other closely, even closer than the thermal wavelength $\lambda_e$ (see Fig.~\ref{fig:spineebad}). When this occurs, a region forms where a big number of electrons and protons can coexist in a single bound state. 

This situation leads to the appearance of non-physical ``clusters'' when the temperature drops below 50 kK, which is approximately the dissociation energy of a hydrogen molecule~\cite{Filinov:PhysRevE:2004, Golubnychiy2004}. We call this behavior as ``collapse into a point'' that should be prohibited by considering Fermi statistics. However, the underlying reasons for this behavior remain unclear, as no significant approximations, aside from~\eqref{eq:dhyfhgdgtg}, were made during the derivation of formula~\eqref{eq:frddtgrf}.

In Ref.~\onlinecite{Filinov:PhysRevE:2004}, it was suggested that the observed effects could be related to many-particle quantum interactions, i.e. the need to take into account all possible permutations of the electron coordinates. This consideration is valid for the conditions considered in Ref.~\onlinecite{Filinov:PhysRevE:2004}, where $\chi \sim 1$. However, in the case of a non-degenerate system ($\chi \ll 1$), the contribution of quantum many-particle effects, i.e. coordinate permutations, to the total interaction should be small.

The failure of explaining clustering in non-degenerate plasma due to the influence of multiparticle quantum interactions becomes clear when analyzing a system consisting of $2+2$ electrons and four protons. In this system, many-particle effects reduce to permutations within each electron subgroup. Since each subgroup contains only two electrons, all interactions can be described as pairwise. Note again that we consider $\chi \ll 1$.
However, even in such a small system at low temperatures, clustering and the collapse of electrons with the same spin projection can still be observed during simulation. 

One possible explanation is that electrons with identical spin projections cannot be treated as point-like particles, as in approximation~\eqref{eq:dhtgrfe}. The finite size of these electrons leads to increased repulsive interaction. This repulsion helps prevent the system's collapse.
The next section will consider a possible approximate solution to address this issue.

\subsection{Fixing the problem of cluster formation with the same electron spin projection}

\begin{figure}[h!]
	\centering
	\includegraphics[width=0.8\linewidth]{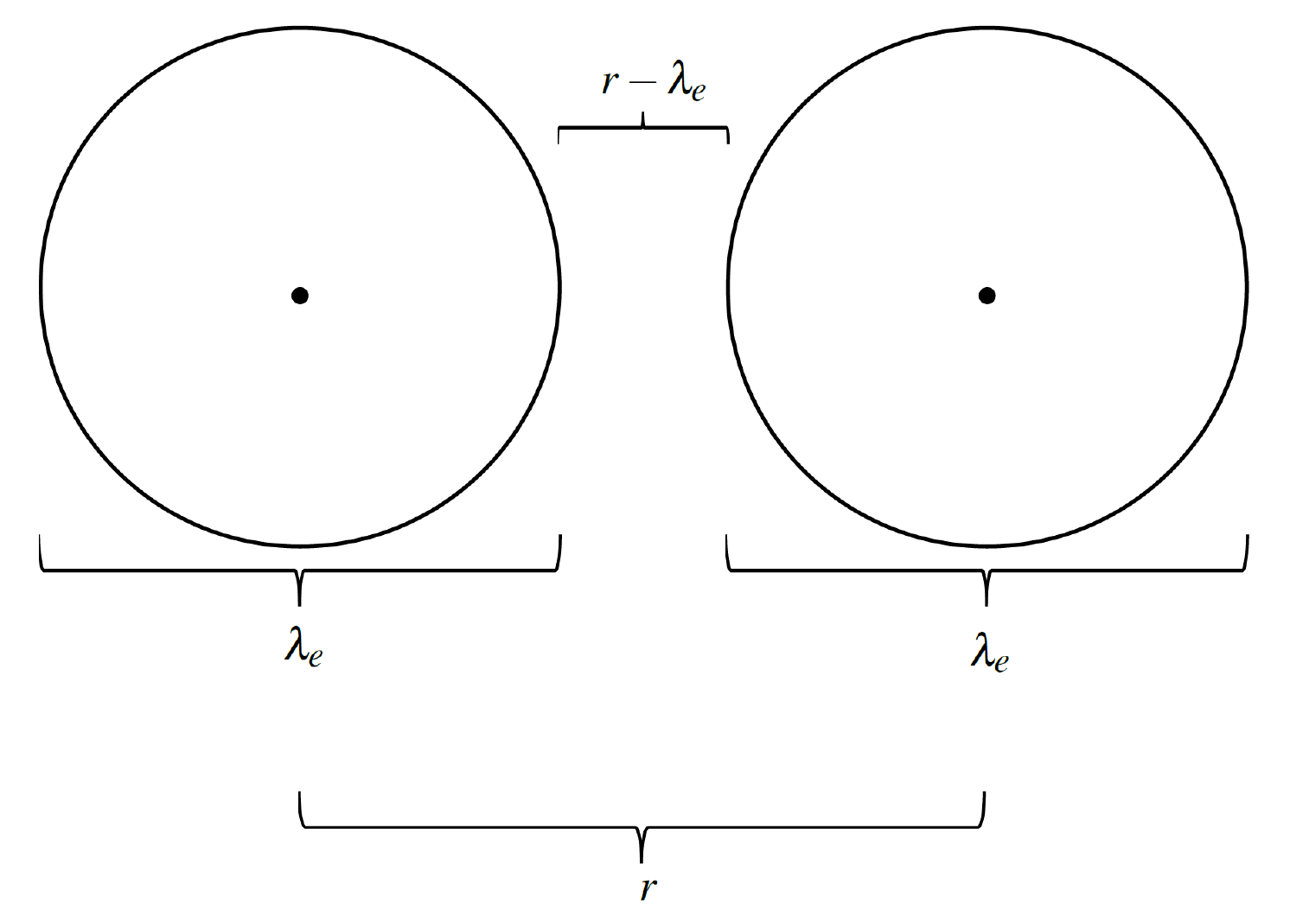}
	\caption{
		A qualitative illustration of how to account for the finite size of two electrons in forces. The characteristic size of two electrons with the same spin projection (identical permutation) are shown. The size of each particle is equal to~$\lambda_{e}$. The centers of these trajectories are located at a distance $r$, while the interaction occurs at smaller distances $\sim (r-\lambda_{e})$. Thus, the repulsion between electrons should increase.}
	\label{fig:twoelectronstrajectories}
\end{figure}

The insufficient repulsion of electrons with the same spin projection can be explained qualitatively as follows. When two electrons with the same spin projection approach each other, their finite size should influence their interaction. A typical size of each electron is related to the reduced thermal  de Broglie wavelength $\lambda_{e} = \Lambda/\sqrt{2\pi}$. Thus, the electrons situated at a distance $r$ from one another interact at a smaller distance of approximately $r - \lambda_{e}$ (see Fig. \ref{fig:twoelectronstrajectories}). 

So, if we consider the finite size of electrons, they should scatter when repelled due to the Pauli principle. In fact, the finite size of electrons is taken into account in the contribution~$\Phi^\text{I}_{0, ee}(\textbf{r}_{12},-\textbf{r}_{12}; \beta)$ of the formula~\eqref{eq:phi_st}. However, the approximation~\eqref{eq:dhtgrfe} is used that no longer takes into account the exchange in potential energy.
Thus, we believe that the problem arises from an approximate treatment of electron Coulomb interaction, probably related to the approximation~\eqref{eq:dhyfhgdgtg}.

Calculation of the exact density matrix, as well as the corresponding exchange contribution $\Phi^\text{I}_{0, ee}(\textbf{r}_{12},-\textbf{r}_{12}; \beta)$ is beyond the scope of this work.
In order to correct the formation of non-physical electron bound states, the following force modification is used. 

In MD, the interelectron interaction force for a triplet state is given by the gradient of p/p:
\begin{equation}
	\label{eq:dghdgsfsef}
	\textbf{f}^\text{T}_{0, ee}(\textbf{r};\beta) = -e^2\nabla  \Phi^\text{T}_{0, ee}({r};\beta).
\end{equation}
This force does not sufficiently repel electrons with the same spin projection. Then, a factor can be introduced into the interaction force that enhances the Coulomb repulsion of electrons at a distance of the order of the de Broglie wavelength. Thus, we account for finite electron size as follows:
\begin{multline}
	\label{eq:fdrgdfsef}
	\textbf{f}^\text{T}_{0, ee}(\textbf{r};\beta) \to -\frac{e^2r^2}{(r - \alpha^\text{T}_{ee} \lambda_{ee})^2}\nabla  \Phi^\text{I}_{0, ee}(\textbf{r},\textbf{r}; \beta)\\+\frac{1}{\beta} \nabla\ln \left(1 - \exp\left[-\frac{m_e r^2}{\hbar^2 \beta}\right]\right).
\end{multline}
Note that the first contribution is responsible for the Coulomb interaction, and the second for the exchange interaction. At $\alpha^\text{T}_{ee} = 0$, this force is equal to~\eqref{eq:dghdgsfsef}.

Thus, the Coulomb repulsion of point electrons works as if the electrons are at a distance $r - \alpha^\text{T}_{ee} \lambda_{ee}$, where $\alpha^\text{T}_{ee}$ is a number of the order of 1. This modification is made only due to the interaction of electrons with the same spin projection; the calculation of the total potential energy, as well as the interaction forces between other particles, remain unchanged. The effect of such a modification of forces can be seen in Fig.~\ref{fig:spineebad}. 

In fact, the value $\alpha^\text{T}_{ee} = 1$ does not always stabilize the system. In this case, the number $\alpha^\text{T}_{ee}$ is chosen to be as small as possible to prevent the formation of non-physical clusters (see Tab.~\ref{tab:alphaGamma}).

Thus, with the help of the described force modification~\eqref{eq:fdrgdfsef}, the final size of electrons is approximately taken into account. Although such a modification is not rigorously justified, it allows the electron subsystem to behave correctly and ensures the formation of molecules without the formation of clusters.

\subsection{Accounting for long-range interactions using the angular-averaged Ewald potential}
If one accounts for long-range interactions with periodic images, the potential part of the Hamiltonian of hydrogen plasma is represented by the following series:
\begin{equation}
	\label{eq:initPotenergy}
	\hat{U}^\text{E} = \cfrac{1}{2} \sideset{}{'}\sum_{\bm{\eta}}\sum_{i,j = 1}^N\cfrac{q_iq_j}{|\textbf{r}_{ij}+L\bm{\eta}|}, \quad \sum_{i=1}^N q_i = 0.
\end{equation}
The prime in summation means that the terms with $\bm{\eta} = \textbf{0}$ are omitted if $i = j$. 
Note that this series is conditionally convergent. To obtain a physically correct result, one should use the Ewald summation technique~\cite{Ewald:1921}. Despite the fact that the resulting expression represents a pair interaction with the Ewald potential~\cite{Demyanov:JChemPhys:2025}, a rather complex expression for the potential does not allow one to solve the Bl\"{o}ch equation analytically using the Kelbg procedure~\cite{Kelbg:1963}. Since the plasma system is isotropic, one can average the Ewald potential over angles and obtain the angular-averaged Ewald potential (AAEP):
\begin{equation}
	\label{eq:AAEP}
	\varphi(r) = 
	\begin{cases}
		\cfrac{1}{r} \left[1 + \frac{r}{2 r_m}((r/r_m)^2 - 3)\right], &r\leq r_m,\\
		0, &r>r_m.
	\end{cases}
\end{equation}
One may find its derivation in Refs.~\onlinecite{Demyanov:JPA:2022, Demyanov:PhysRevE:2022}. As a result, we obtain the interaction potential of finite radius $r_m = (4\pi/3)^{-1/3}L$, and the potential part of the Hamiltonian  is written as follows:
\begin{equation}
	\label{eq:classical_energy_pot}
	\hat{U}^\text{a} = U_0 + \cfrac{1}{2}\sum_{i = 1}^N\sum_{\substack{j\in \mathcal{S}(\textbf{r}_{i}) \\ i \neq j}}q_iq_j\varphi(r_{ij}),\quad U_0 = -\cfrac{3 e^2N}{4 r_m},
\end{equation}
where $\mathcal{S}(\textbf{r}_{i})$ denotes a sphere of radius $r_m$ centered at $\textbf{r}_{i}$~\cite{Yakub:JChemPhys:2003, Jha:2010, Demyanov:JPA:2022}. In this way, the minimum image convention is used~\cite{Demyanov:JPA:2022, Brush:JChemPhys:1966}.

Substituting AAEP into Eq.~\eqref{eq:diag_kelbg_gen}, one can obtain a contribution to the Kelbg p/p that takes into account the Coulomb long-range interaction~\cite{Demyanov:CPP:2022, Demyanov:CPC:2024}:
\begin{equation}
	\label{eq:diagkelbg}
	\Phi(\textbf{r},\textbf{r}; \beta) \equiv \Phi(r; \beta) = \Phi_0(r; \beta) + \Phi_1(r; \beta),
\end{equation}
where
\begin{equation}
	\label{eq:phi1Def}
	\Phi_1(r;\beta) =  \frac{4}{r\pi} 
	\left[
	I_{\text{all}}\left(\frac{r}{\lambda}, \frac{r_m}{\lambda}\right) -I_{\text{all}}\left(-\frac{r}{\lambda}, \frac{r_m}{\lambda}\right) -\frac{r}{4\lambda} \pi ^{3/2}
	\right],
\end{equation}
and
\begin{equation}
	I_{\text{all}}(x, x_m) = I_{\text{exp}}(x, x_m) + I_{\erf}(x, x_m) + I_{\text{mod}}(x, x_m),
\end{equation}
\begin{multline}
	\label{eq:expcontr}
	I_{\text{exp}}(x, x_m) = \frac{\pi  e^{-(x_m+x)^2} (x_m+x)}{128 x_m^3 |x_m+x|} (2 x^2 x_m-2 x^3+10 x x_m^2\\-5 x+6 x_m^3+3
	x_m),
\end{multline}
\begin{multline}
	I_{\erf}(x, x_m) = \frac{\pi ^{3/2}}{256 x_m^3} (4 (x_m+x) ((x^2+3) x_m-x (x^2+3)\\+5 x x_m^2+3
	x_m^3)-3) \erf\left(|x_m+x|\right),
\end{multline}
\begin{equation}
	\label{eq:modcontr}
	I_{\text{mod}}(x, x_m) = \frac{\pi  |x_m+x|}{16 x_m^3 (x_m+x)} \left(x^3-3 x x_m^2+x-2 x_m^3\right).
\end{equation}
The derivation of Eq.~\eqref{eq:diagkelbg} can be found in Ref.~\onlinecite{Demyanov:CPP:2022}. We call Eq.~\eqref{eq:diagkelbg} as Kelbg-AAE p/p.

One of the major properties of the Kelbg-AAE p/p~\eqref{eq:diagkelbg} is that it tends to zero at large distances:
\begin{equation}
	\label{eq:fdgthsefe}
	\lim_{r\to\infty}\left(\exp\left(r^{(2-\xi)}\right)\Phi(r; \beta)\right) = 0, \quad \forall \xi~>~0.
\end{equation}
This behavior is ensured by the fact that at large distances $r_{ij}\gg \lambda_{ij}$ the p/p $\Phi(\textbf{r}_{ij},\textbf{r}_{ij}; \beta)$ in Eq.~\eqref{eq:diag_kelbg_gen} mostly coincides with the classical interaction potential $\phi(\textbf{r}_{ij})$ that corresponds to it. Since the Kelbg-AAE p/p was obtained on the basis of the AAEP that is equal to zero at $r\geq r_m$, this property is also preserved for the~p/p.

Thus we define the improved Kelbg-AAE p/p as following~\cite{Demyanov:CPC:2024}:
\begin{equation}
	\label{eq:diagKelbgAAEPPImroved}
\Phi^\text{I}(r_{ij}; \beta) = \Phi^\text{I}_0(r_{ij};\beta) + \Phi_1(r_{ij}; \beta).
\end{equation}
The contribution $\Phi_1(r_{ij}; \beta)$ can be included and excluded in the calculation procedure, investigating its influence on the rate of convergence of thermodynamic properties in terms of the number of particles~\cite{Demyanov:CPP:2024}. Note that the interaction radius is set equal to $r_m$ both in the case of taking into account the contribution of $\Phi_1(r_{ij}; \beta)$ and without it. Note that $r_m>L/2$, which may cause some peculiarities in a simulation (for details one may see Refs.~\onlinecite{Jha:2010, Demyanov:JPA:2022, dornheim2025applicationsphericallyaveragedpair}).

\subsection{Calculation of forces, energy and pressure}
To perform MD simulation, in addition to pseudopotentials, it is necessary to know interparticle forces to solve the equations of motion. Thus, the forces between electrons and protons, $\textbf{f}_{ep}$, two electrons, $\textbf{f}^\text{S(T)}_{ee}$, and two protons, $\textbf{f}_{pp}$, are given by the following equations:
\begin{align}
	\textbf{f}_{ep} &= e^2\nabla \Phi^\text{I}_{ep}({r}; \beta), \label{eq:forceep}\\
	\textbf{f}^\text{S}_{ee} &= -e^2\nabla \Phi^\text{S}_{ee}({r}; \beta),\\
	\textbf{f}^\text{T}_{ee} &= -\frac{e^2r^2}{(r - \alpha^\text{T}_{ee} \lambda_{ee})^2}\nabla  \Phi^\text{I}_{ee}({r}; \beta)+\frac{1}{\beta} \nabla\ln (1 - e^{-{r^2}/{\lambda_e^2}}),\\
	\textbf{f}_{pp} &= -e^2\nabla \Phi({r}; \beta), \label{eq:forcepp}
\end{align}
where the improved Kelbg-AAE p/p is defined in Eq.~\eqref{eq:diagKelbgAAEPPImroved}. Recall also that the parameter $\alpha_{ee}^\mathrm{T}$ was introduced into the electron-electron interaction forces for electrons with the same spin projection to take into account the finite size of electrons. The dependence of $\alpha_{ee}^\mathrm{T}$ on the coupling parameter is indicated in the Table~\ref{tab:alphaGamma}.

As already mentioned, the interaction radius is always equal to $r_m$. Since the improved Kelbg p/p is not equal to zero at $r=r_m$, all forces~\eqref{eq:forceep}--\eqref{eq:forcepp} are always shifted by its value at $r=r_m$ to avoid numerical effects at the boundary of the sphere~\cite{rapaport_2004}. For convenience, the same shifting is applied in the case of Eq.~\eqref{eq:diagKelbgAAEPPImroved}, though its effect is almost negligible due to~\eqref{eq:fdgthsefe}.

\begin{table}[h!]
	\centering
	\caption{Selected values of the parameter $\alpha^\text{T}_{ee}$ in the interaction forces between electrons with the same spin projection at $\chi = 0.01$. }
	\label{tab:alphaGamma}
	\begin{tabular}{|c|c|c|c|c|c|c|c|c|c|c|c|}
		\hline
		$\Gamma$               & 0.1 & 0.25 & 0.4& 0.5 & 0.65& 0.8 & 1    & 1.5 & 2 & 2.5 & 3 \rule{-2pt}{2pt}\\ \hline
		$\alpha^\text{T}_{ee}$ & 0   & 0 &  1 & 1 & 1.25 & 1.5 & 1.75 & 2.25   & 3 & 4   & 5 \rule{-2pt}{9pt}\\ \hline
	\end{tabular}
\end{table}

The main quantity that determines the probability of some configuration in the $NVT$-ensemble is action~\cite{Demyanov:CPC:2024, RevModPhys.67.279}: 
\begin{equation}
	\label{eq:actionPhiH}
S(\textbf{R}, \beta) = \beta U_0 + \frac{\beta}{2}\sum_{i = 1}^N\sum_{\substack{j\in \mathcal{S}(\textbf{r}_{i}) \\ i \neq j}}q_i q_j \Phi_{ij}(r_{ij};\beta),
\end{equation}
where $\Phi_{ij}(r_{ij};\beta)$ depends on the particle type.  For electron-proton interaction $\Phi_{ij}(r_{ij};\beta)= \Phi^\text{I}_{ep}({r}_{ij}; \beta)$, for electron-electron interaction in a singlet state $\Phi_{ij}(r_{ij};\beta)= \Phi^\text{S}_{ee}({r}_{ij}; \beta)$, for electron-electron interaction in a triplet state $\Phi_{ij}(r_{ij};\beta)= \Phi^\text{T}_{ee}({r}_{ij}; \beta)$, and for proton-proton interaction $\Phi_{ij}(r_{ij};\beta)= \Phi({r}_{ij}; \beta)$.

First, we find a formula for calculating the pressure of non-degenerate hydrogen plasma. Note that the virial pressure from LAMMPS gives the correct value only in the case of not accounting for long-range interaction~\cite{Onegin_2024, Demyanov:JPhysA:2025}. To obtain correct pressure values, we can use formula (4) in Ref.~\onlinecite{Louwerse2006} (see also Ref.~\onlinecite{Onegin_2024}), replacing the potential energy with the action~\eqref{eq:actionPhiH}:
\begin{equation}
	\beta P(\textbf{R})
	=N/V - \left(\frac{\partial (\beta S(\textbf{R}; \beta))}{\partial V}\right)_T.
\end{equation}
Since the contributions from different types of particles are not the same, it is convenient to introduce the following form:
\begin{equation}
	\label{eq:ggdrfeafae}
	\beta P(\textbf{R})V = N + \frac{\beta U_0}{3} + \beta\sum_{i = 1}^N\sum_{\substack{j\in \mathcal{S}(\textbf{r}_{i}) \\ i \neq j}}q_i q_j p_{ij}({r}_{ij}, r_m, \beta),
\end{equation}
where the two-particle pressure contribution is:
\begin{multline}
	\label{eq:gdtgftgrfs}
	p_{ij}({r}_{ij}, r_m, \beta) \\= p^\text{I}_{0}({r}_{ij}, \beta)+p_{1}({r}_{ij}, r_m, \beta)+\delta_{i,e}\delta_{j,e}p^\text{S(T)}_{ij}({r}_{ij}, \beta).
\end{multline}

For convenience, we also introduce the notations $x_{ij}~=~r_{ij}/\lambda_{ij}$ and $x_m = r_m/\lambda_{ij}$. To calculate the volume derivative we use volume and coordinate scaling~\cite{Onegin_2024}, $V\to \nu^3 V$, $\textbf{R}\to \nu \textbf{R}$, $r_m\to \nu r_m$, and $\partial/\partial V \to (3V)^{-1}\partial/\partial\nu$.
For the contribution from the improved Kelbg p/p, we obtain:
\begin{multline}
	p^\text{I}_{0}({r}_{ij}, \beta) = -\frac{1}{6}\left.\frac{\partial}{\partial \nu} \Phi^\text{I}_0(\nu {r}_{ij}, \beta)\right|_{\nu = 1}
	\\=
	\frac{1}{6r_{ij}}\left[1 + 2x_{ij}^2e^{-\gamma_{ij}x^2_{ij}}-e^{-x^2_{ij}}(1+2x^2_{ij})\right].
\end{multline}

The contribution responsible for the long-range interaction (for $r_{ij}\leq r_m$) has the following form:
\begin{multline}
	p_{1}({r}_{ij}, r_m, \beta) = -\frac{1}{6}\left.\frac{\partial}{\partial \nu} \Phi_1(\nu {r}_{ij}, \nu r_m, \beta)\right|_{\nu = 1}\\
	=
	-\frac{1}{96 r_{ij} x_m^3}\Big[3 \sqrt{\pi } \left(2 x_{ij}^2-2 x_m^2+1\right) (\erf(x_{ij}-x_m)\\+\erf(x_{ij}+x_m))-8 x_{ij} \left(x_{ij}^2-3 x_m^2+3\right)\\+6 e^{-(x_{ij}+x_m)^2} (x_{ij}-x_m)+6
	e^{-(x_{ij}-x_m)^2} (x_{ij}+x_m)\Big].
\end{multline}
If the long-range interaction is not taken into account, then the contributions $\beta U_0/3$ and $p_{1}({r}_{ij}, r_m, \beta)$ are equal to zero.

The additional contribution responsible for the opposite projections of electron spins has the form:
\begin{equation}
	p^\text{S(T)}_{ij}({r}_{ij}, \beta)
	=
	-\frac{1}{\beta e^2}\frac{x_{ij}^2}{1\pm e^{-x_{ij}^2}}.
\end{equation}
In this case, the notation $\delta_{i,e}\delta_{j,e}$ in Eq.~\eqref{eq:gdtgftgrfs} means that this contribution is non-zero only in the case when particles $i$ and $j$ are electrons. 

Next, we write out the formula for calculating the total energy:
\begin{equation}
	\label{eq:fullPotEnergyQuasiClMD}
	\beta E(\textbf{R}) = \frac{3N}{2}+\frac{\Gamma}{2}
	\sum_{i = 1}^N\sum_{\substack{j = 1\\ i\neq j}}^{N_{s,i}} r_a\mathcal{F}_{ij}({r}_{ij}; r_m, \beta)
	\\-  \frac{3N\Gamma}{4r_{m}/r_a},
\end{equation}
where
\begin{multline}
	\label{eq:Fmathcalij}
	\mathcal{F}_{ij}({r}_{ij}; r_m, \beta)
	\\=
	\mathcal{F}^\text{I}_{0}(r_{ij}, \beta)
	+
	\mathcal{F}_{1}(r_{ij}; r_m, \beta)
	+
	\delta_{i,e}\delta_{j,e}\mathcal{F}^\text{S(T)}_{ij}({r}_{ij}, \beta)
\end{multline}
has the same form as Eq.~\eqref{eq:gdtgftgrfs}. 

Then the contribution from the improved Kelbg p/p has the form:
\begin{multline}
	\mathcal{F}^\text{I}_{0}(r_{ij}, \beta)
	=
	\Phi^\text{I}_0({r}_{ij}, \beta) + \beta \left(\frac{\partial \Phi^\text{I}_0({r}_{ij}, \beta)}{\partial \beta}\right)_V
	\\\equiv
	\Phi^\text{I}_0({r}_{ij}, \beta) + \beta \left(\frac{\partial \Phi^\text{I}_0({r}_{ij}, \lambda_{ij}(\beta), \gamma_{ij}(\beta))}{\partial \beta}\right)_V.
\end{multline}
The temperature derivative of the improved Kelbg p/p is derived in App.~\ref{app:1} (see equation~\eqref{eq:frbdgrsvgdbg}). 

The contribution responsible for the long-range interaction has the following form (see equation~\eqref{eq:phi1Def}):
\begin{equation}
	\mathcal{F}_{1}(r_{ij}; r_m, \beta)
	=
	\Phi_1({r}_{ij};\beta) + \beta \left(\frac{\partial \Phi_1({r}_{ij};\beta)}{\partial \beta}\right)_V.
\end{equation}
The temperature derivative of the 
additional term $\Phi_1({r}_{ij};\beta)$
is derived in App.~\ref{app:1} (see equation~\eqref{eq:rdgdgfrf}). 

The additional contribution responsible for the opposite projections of electron spins has the following form:
\begin{multline}
	\mathcal{F}^\text{S(T)}_{ij}({r}_{ij}, \beta)
	=
	-\frac{1}{\beta e^2} \beta \left(\frac{\partial}{\partial \beta}\ln \left(1 \pm \exp\left[-\frac{m_e r^2}{\hbar^2 \beta}\right]\right)\right)_V \\=
	\frac{1}{\beta e^2}
	\frac{\mp \exp\left[-\frac{r^2}{\lambda_{e}^2}\right]\frac{r^2}{\lambda_{e}^2}}{\left(1 \pm \exp\left[-\frac{r^2}{\lambda_{e}^2}\right]\right)}.
\end{multline}

Eq.~\eqref{eq:ggdrfeafae}~and~\eqref{eq:fullPotEnergyQuasiClMD} will be used to calculate the pressure and energy of MD configurations in LAMMPS simulations. Interparticle forces in MD are given by formulas~\mbox{\eqref{eq:forceep}--\eqref{eq:forcepp}}.

Note that the forces and contributions to the potential energy are inconsistent  with each other, as in the case of a true classical system. This is due to the fact that the interaction forces must correspond to the action in which there are no additional temperature contributions $(\partial \Phi^\text{I}(r; \beta)/\partial \beta)_V$ associated with the explicit dependence of the p/p on temperature. 

\section{Simulation parameters and details~\label{sec:sim}}

\subsection{Simulation parameters}
A dimensionless system of units is used for MD simulations in LAMMPS: energy is measured in units of $\varepsilon$, length is measured in units of $\sigma$, and mass in units of $m$, which are defined as follows:
\begin{equation}
	\label{eq:unitsLammpsH}
	\varepsilon = e^2/r_a, \quad \sigma = r_a, \quad m = m_e.
\end{equation}
In addition, it is convenient to recall the atomic units of time $t_\text{at} = \hbar/E_H$ to relate real time to the time in LAMMPS. For example, the time step $\Delta t^\text{L}$ specified in the input file of a simulation can be related to the time step $\Delta t$ measured in seconds as follows:
\begin{equation}
	\Delta t = \Delta t^\text{L} \sqrt{\frac{m \sigma^2}{\varepsilon}} = t_\text{at} \Delta t^\text{L} \sqrt{\frac{9\pi \Gamma^3}{2\chi^2}}.
\end{equation}
Any simulation time interval and the one measured in seconds are related by the coefficient $t_\text{at} \sqrt{9\pi \Gamma^3/(2\chi^2)}$.

In addition, plasma exhibits a characteristic relaxation time associated with its internal processes, which corresponds to the period of plasma oscillations. This parameter can be defined separately for electrons and protons. To ensure accurate simulation results, the total simulation duration should significantly exceed the plasma oscillations period of the heaviest particles (in this case, protons). Accordingly, the simulation time step $\Delta t^{\text{L}}$ is selected to maintain the system's temperature in the $NVT$-ensemble preventing substantial jumps in other thermodynamic parameters, such as total energy, during the simulation.

\begin{figure*}[ht!]
	\centering
	\begin{subfigure}[t]{0.49\linewidth}
		\includegraphics[width=\linewidth]{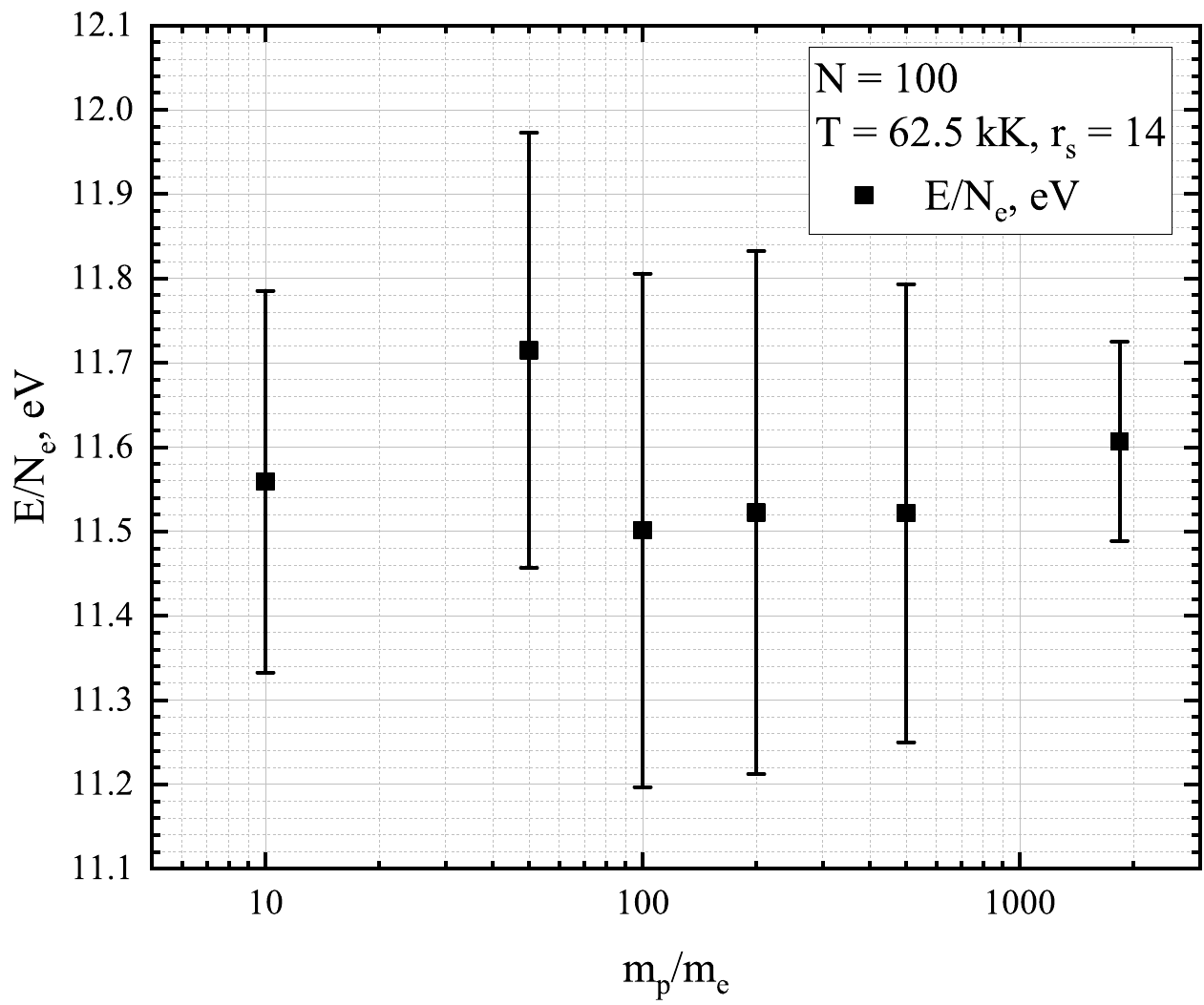}
		\label{fig:massratio1}
	\end{subfigure}
	\hfill
	\begin{subfigure}[t]{0.49\linewidth}
		\includegraphics[width=\linewidth]{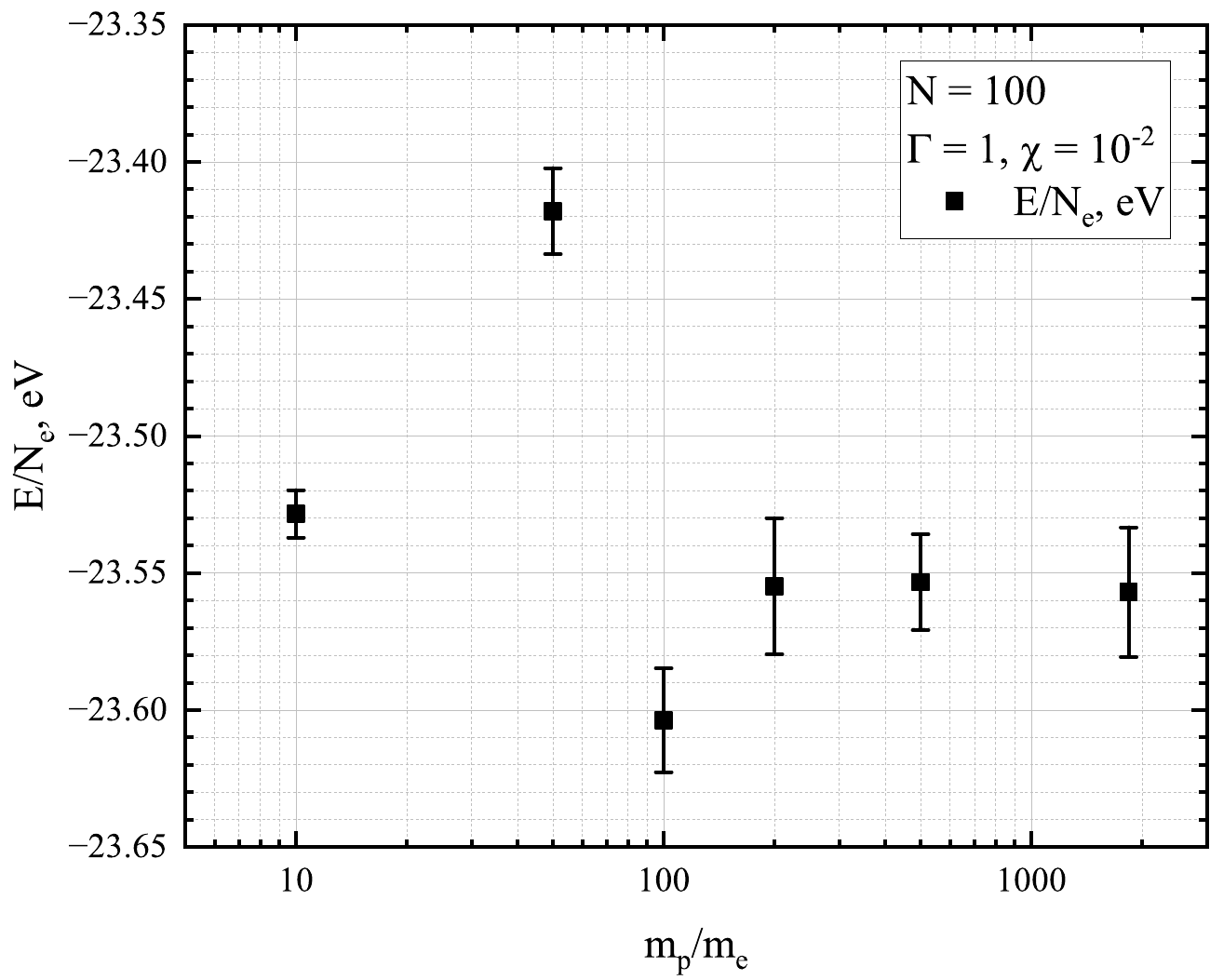}
		\label{fig:massratio2}
	\end{subfigure}
	\caption{Dependence of the total energy of hydrogen plasma on the mass ratio \(m_p/m_e\) in MD simulations under two conditions: (left) absence of molecules and (right) presence of molecules.}
	\label{fig:massratio}
\end{figure*}

We now express the plasma oscillation period in atomic time units using the plasma parameters $\Gamma$ and $\chi$. The plasma frequency for electrons and corresponding oscillation period are given by:
\begin{equation}
\omega_e = \sqrt{\frac{4\pi n_e e^2}{m_e}}, \quad \tau_e = \frac{2\pi}{\omega_e}.
\end{equation}
Now substituting the electron density from Eq.~\eqref{eq:GammaChiDef}, we obtain the relationship between the period of plasma oscillations and plasma parameters:
\begin{equation}
	\tau_e = t_{at} \sqrt{6\pi^3\Gamma^3}/\chi, \quad \tau_p = \sqrt{m_p/m_e}\tau_e.
\end{equation}
The plasma oscillation period for protons is $\sqrt{m_p/m_e}$ times larger than for electrons. Note that the mass ratio in simulations differs from the real one (see Fig.~\ref{fig:massratio}; the ratio $m_p/m_e = 200$ is used in typical MD simulations).

As a result, if $m_\text{tot}$ steps are performed during the simulation, the total simulation time relates to the period of electron oscillations as follows:
\begin{equation}
	\frac{m_\text{tot} \Delta t}{\tau_e} = \frac{\sqrt{3}}{2\pi} m_\text{tot} \Delta t^\text{L}, \quad \omega_e \Delta t = \sqrt{3} \Delta t^\text{L}.
\end{equation}
For the proton oscillation period, the relation is:
\begin{equation}
	\label{eq:dgdtbdtdrf}
	\frac{m_\text{tot} \Delta t}{\tau_p} = \sqrt{\frac{m_e}{m_p}} \frac{\sqrt{3}}{2\pi} m_\text{tot} \Delta t^\text{L}, \quad \omega_p \Delta t = \sqrt{3 \frac{m_e}{m_p}} \Delta t^\text{L}.
\end{equation}
Both of these values should be much greater than one. Since the condition involving the proton oscillation period is stronger, the following inequality should be satisfied:
\begin{equation}
	\label{eq:gdbdfsfes}
	\sqrt{\frac{m_e}{m_p}}\frac{\sqrt{3}}{2\pi} m_\text{tot}\Delta t^\text{L} \gg 1\text{ \ or \ }
	m_\text{tot} \gg \frac{2\pi}{\sqrt{3} \Delta t^\text{L}} \sqrt{\frac{m_p}{m_e}}.
\end{equation}
Therefore, the number of simulation steps should be chosen to be at least on the order of $20 \pi \sqrt{m_p/m_e} / (\sqrt{3} \Delta t^\text{L})$.

Thus, the number of steps is inversely proportional to the size of the time step. The non-degenerate regime assumes a small degeneracy parameter, $\chi \ll 1$. However, choosing a very small $\chi$ requires a correspondingly small time step in MD simulations because the electron orbital speed increases significantly. This effect greatly increases the amount of computations required. We will demonstrate this below.

The characteristic speed of electron orbital motion in a hydrogen atom orbit is given by $v_e = \hbar /(m_e a_B)$. In  simulations, the unit of velocity is defined as $\sqrt{\varepsilon/m}$~\eqref{eq:unitsLammpsH}. Therefore, the dimensionless electron rotation velocity can be expressed as:
\begin{equation}
	v_e \sqrt{m/\varepsilon} = \frac{\hbar}{m_e a_B} \times \sqrt{\frac{e^2}{m_e r_a}} = \sqrt{r_s} \propto \frac{1}{\chi^{1/3}}.
\end{equation}
Thus, as the degeneracy parameter decreases, the dimensionless electron rotation velocity increases. Consequently, a smaller simulation time step is required to accurately capture the sharper rotation trajectory. This sets a lower limit on the degeneracy parameter value. In this study, we use $\chi = 10^{-2}$. This value is small enough to neglect exchange effects and also to choose a sufficiently large step to satisfy condition~\eqref{eq:gdbdfsfes} at $m_\text{tot} = 10^7$.

Equation~\eqref{eq:gdbdfsfes} states that the required number of simulation steps is proportional to the square root of the proton-to-electron mass ratio. The reason is that protons, being much heavier, move significantly slower than electrons. To increase computational efficiency, a reduced proton-to-electron mass ratio is used while ensuring that key physical properties remain unchanged. For example, in Ref.~\onlinecite{Lavrinenko:CPP:2024}, a reduced mass ratio of 200 is employed, which is approximately ten times smaller than the actual physical value.

To investigate the effect of the mass ratio $m_p/m_e$ on the simulation results, this ratio was varied from the real value to 10, and the total energy of hydrogen plasma was calculated. Two temperature regimes were considered. First, the temperature was set to be above 52~kK but below 150~kK; in this case, free particles and atoms coexist. Second, the temperature was set below 52~kK; in this case, molecules are present. This variety is important because at lower temperatures, molecular formation strengthens the sensitivity of simulation results to deviations in proton mass.

Figure~\ref{fig:massratio} illustrates the total energy per electron of hydrogen plasma as a function of the mass ratio. It is evident that at high temperatures (\(T~=~62.5\) kK), the energy dependence on the mass ratio is negligible. In contrast, at lower temperatures and for \(m_p/m_e \leq 100\), the behavior becomes non-monotonic, with significant deviations from the results obtained using the real mass ratio. Therefore, an optimal value of \( m_p/m_e = 200 \) was chosen for the simulations presented below.

Finally, we use the following simulation parameters. Step size is $\Delta t^\text{L}~=~5\times 10^{-4}$. As the equilibrium is reached, $m_\text{tot}~=~10^7$ steps of MD are performed. Additionally, we set $m_p/m_e~=~200$ for all MD simulations. Note that the calculation of thermodynamic functions, forces, and pseudopotentials was performed for the real mass ratio. Statistical errors are calculated using the block averaging technique, so the whole equilibrium section of the simulation is divided into 5~blocks~\cite{Demyanov:PhysRevE:2022}. Interparticle forces and energy contributions~\eqref{eq:Fmathcalij} are calculated using KelbgLIP code~\cite{Demyanov:CPC:2024} on a uniform grid for the variable $u = r^2$ with a step $\Delta u =10^{-4}a^2_B$.

\subsection{Identification of atoms, molecules, and ions and ionization degree of hydrogen plasma \label{sec:algcomplex}}

In numerical simulations in which all particles are represented by points, it is possible to distinguish between free particles and various bound states, such as atoms, molecules, and charged complexes, on a microscopic level. This feature enables a direct determination of plasma composition and ionization degree, defined as the fraction of free electrons~\cite{Filinov:PhysRevE:2023}.

To determine the composition, we employ a cluster analysis based on interparticle distances. If the distance between a pair of particles is smaller than an adjusted threshold, those particles are assigned to the same cluster. Specifically, we introduce a threshold parameter, $d_{H}$, for the electron-proton distance and $d_{HH}$ for the proton-proton distance. These values are determined from the analysis of radial distribution functions (RDFs) and serve as input parameters of the algorithm described below.

\begin{figure}[h!]
	\centering
	\includegraphics[width=1.0\linewidth]{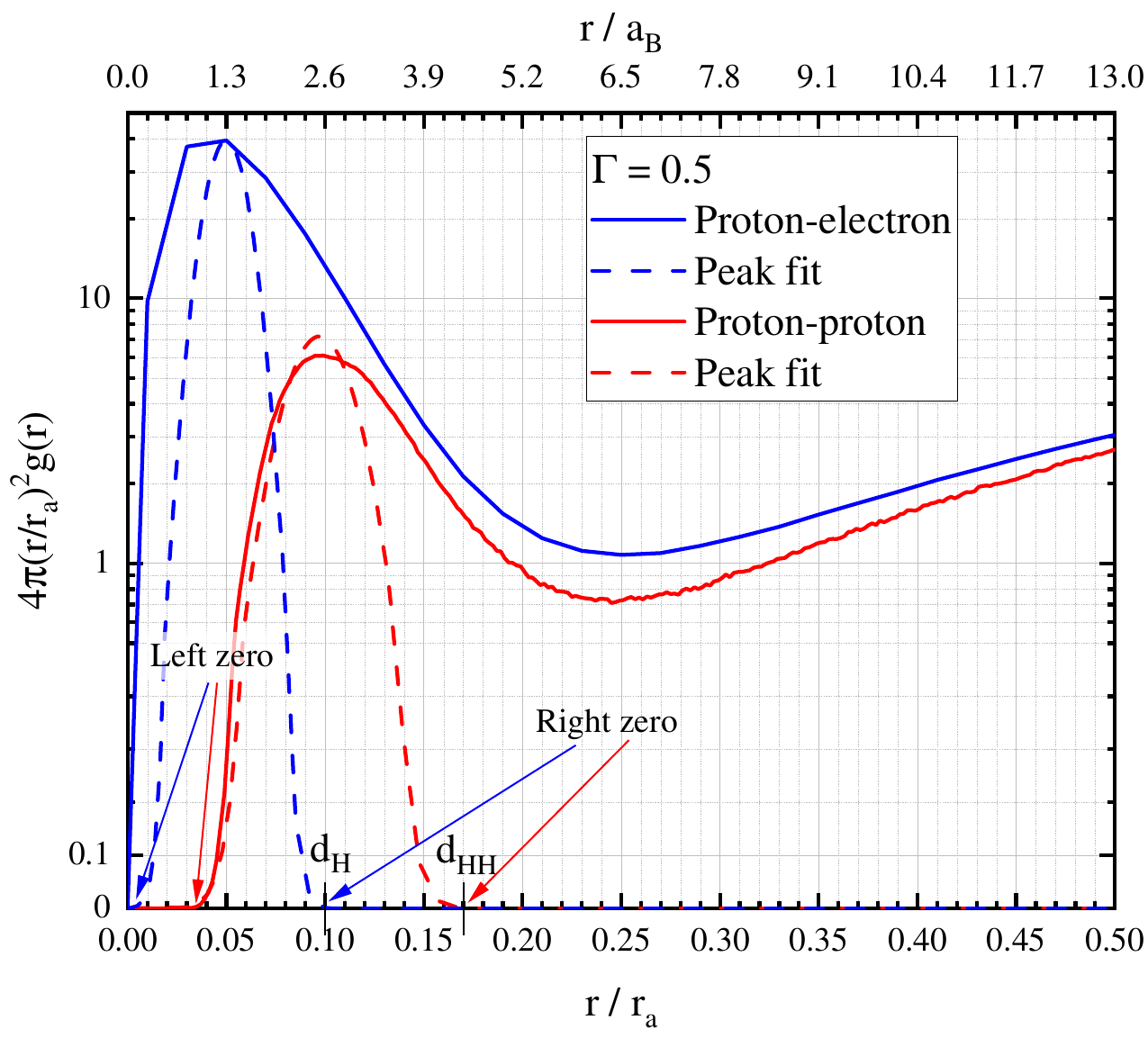}
	\caption{Determination of interparticle distance threshold value: $4\pi r^2 g(r)$ for electron-proton (blue) and proton-proton (red) pairs. Dashed vertical lines illustrate the peak and half-width construction.}
	\label{fig:rhrhhfig}
\end{figure}

The values of $d_H$ and $d_{HH}$ are extracted from the behavior of the RDFs, i.e., $g_{p-e}(r)$ for proton-electron and $g_{p-p}(r)$ for proton-proton. By considering the quantity $4\pi r^2 g(r)$, representing the probability density to find a particle at a distance $r$, we identify characteristic peaks corresponding to bound states (Fig.~\ref{fig:rhrhhfig}). For each pair type, the position of the peak and the location of the left-sided minimum (``left zero'') are determined; the half-width is defined as the distance between these points. The standard threshold is then set as twice the half-width added to the left zero. In systems with pronounced peaks and rapid decay (as presented at high coupling), $d_H$ or $d_{HH}$ may instead be defined by the location of the right zero, i.e., the point where $4\pi r^2 g(r)$ vanishes beyond the peak.

The following species are considered in the cluster analysis algorithm:
\begin{itemize}
	\item $H^+$ : free (not bound) proton
	\item $H$ : neutral atom (one electron + one proton)
	\item $H^-$ : negative ion (one proton + two electrons)
	\item $H_2$ : hydrogen molecule (two protons + two electrons)
	\item $H_2^+$ : molecular ion (two protons + one electron)
	\item $H_3^+$ : triatomic ion (three protons + two electrons)
	\item $H_3^{2+}$ : triatomic dication (three protons + one electron)
\end{itemize}
Any electrons not found in these clusters are tagged as free (not bound).

The algorithm for determining the composition of the system consists of several stages. Initially, there is a list of all protons and electrons.
In the first step, the nearest electron is found for each proton. If the distance between them is less than $d_H$, they are combined into an $H$ atom. Next, both particles are excluded from the list of available protons and electrons. As a result, the list of ``primary'' atoms is formed, as well as the lists of the remaining free protons and electrons (these lists can be empty).

In the second step, the obtained primary atoms are considered. Pairs of atoms whose protons are at a distance less than $d_{HH}$ are combined into $H_2$ molecules. The used atoms are excluded from the list of primary ones. As a result, the list of primary molecules and the list of the remaining primary atoms are formed.

Next, a search for $H_2^+$ complexes is performed. To achieve this, among the remaining primary atoms and free protons, such pairs are found in which the distance between the protons is less than $d_{HH}$. The combinations found are combined into $H_2^+$ complexes that are also excluded from further consideration.

The $H_3^+$ and $H_3^{2+}$ complexes are determined in a similar manner. To find $H_3^+$, $H_2$ molecules and free protons are considered. If the distance between one of the molecule's protons and a free proton is less than $d_{HH}$, the particles are combined into an $H_3^+$ complex. When determining $H_3^{2+}$, the distance between the protons of the $H_2^+$ complexes and the remaining free protons is checked. At each stage, all found compounds are excluded from the list of available ones.

The final step is to detect the negatively charged $H^-$ ions. To do this, the distance between the remaining atoms and free electrons is checked: if it is less than $d_H$, the $H^-$ complex is formed. Protons and electrons that are not included in the complexes are considered as free.

The fraction of each complex is given as the ratio of the number of protons contained in that complexes to the total number of protons in the system. For example, $N_{H_2}$ molecules contain $2 N_{H_2}$ protons, so the molecular fraction is $2N_{H_2}/N_p$. All fractions of considered complexes sum to unity. The degree of ionization is calculated as the number of free electrons divided by the total number of electrons. These values are averaged over simulation snapshots to obtain equilibrium values.

The algorithm is implemented in Python. Efficient neighbor search utilizes a $k$-$d$-tree ($k$-dimensional tree as provided by \texttt{scipy.spatial}), and periodic boundary conditions are included using minimum image convention.

\section{Results and discussion \label{sec:results}}

\subsection{Verification on path integral Monte Carlo simulations}

\begin{table*}
\centering
\begin{minipage}{\textwidth}
	\centering
	\caption{Energy and pressure values for the isotherm $T=62.5$~kK. In the MD simulation, $N_e = 34$ was used. For comparison, data from Ref.~\onlinecite{Filinov:PhysRevE:2023}, obtained by the PIMC method, are given. The differences are calculated as follows: $\delta_E = |E^\text{MD}/E^\text{PIMC}-1|\times100\%,
		\delta_P = |P^\text{MD}/P^\text{PIMC}-1|\times100\%$.}
	\label{tab:62500H}
	\begin{tabular}{|l|ll|ll|ll|ll|}
		\hline
		$r_s$ & $\chi \times 10^2$ & $\Gamma$ &
		$E^\text{PIMC}$, eV & $P^\text{PIMC}$, kbar &
		$E^\text{MD}$, eV & $\delta_E$,\% &
		$P^\text{MD}$, kbar & $\delta_P$,\% \\ \hline
		10 & 4.27 & 0.51 & 9.450 & 23.3774 & 8.5(5) & 9.82 & 23.00(7) & $1.62$ \\
		14 & 1.56 & 0.36 & 11.954 & 9.0889 & 11.6(2) & 2.61 & 9.01(2) & $0.87$ \\ \hline
		20 & 0.53 & 0.25 & 13.787 & 3.267 & 13.8(2) & 0.33 & 3.258(6) & $0.28$ \\
		40 & 0.07 & 0.13 & 15.446 & 0.42666& 15.4(2) & 0.56 & 0.4251(7) & $0.37$ \\
		60 & 0.02 & 0.08 & 15.802 & 0.127664& 15.73(3)& 0.47 & 0.12719(4)& $0.37$ \\ \hline
	\end{tabular}
\end{minipage}

\begin{minipage}{\textwidth}
	\centering
	\caption{Energy and pressure values for the isotherm $T=50$~kK (see caption to Table~\ref{tab:62500H}).}
	\label{tab:50000H}
	\begin{tabular}{|l|ll|ll|ll|ll|}
		\hline
		$r_s$ & $\chi \times 10^2$ & $\Gamma$ & $E^\text{PIMC}$, eV & $P^\text{PIMC}$, kbar & $E^\text{MD}$, eV & $\delta_E$, \% & $P^\text{MD}$, kbar & $\delta_P$, \% \\ \hline
		17  & 1.21 & 0.37 & 8.561  & 3.9738     & 8.0(4)     & 6.6 & 3.92(2)      &  1.4 \\ \hline
		25  & 0.38 & 0.25 & 10.628 & 1.32898    & 10.3(4)    & 3.2 & 1.315(5)     &  1.1 \\
		50  & 0.05 & 0.13 & 12.284 & 0.174537   & 12.20(11)  & 0.7 & 0.17373(11)  &  0.5 \\
		100 & 0.01 & 0.06 & 12.751 & 0.022157   & 12.68(2)   & 0.5 & 0.0220620(7) &  0.4 \\ \hline
	\end{tabular}
\end{minipage}

\begin{minipage}{\textwidth}
	\centering
	\caption{Energy and pressure values for the isotherm $T=31250$~K (see caption to Table~\ref{tab:62500H}).}
	\label{tab:31250H}
	\begin{tabular}{|l|ll|ll|ll|ll|}
		\hline
		$r_s$ & $\chi \times 10^2$ & $\Gamma$ & $E^\text{PIMC}$, eV & $P^\text{PIMC}$, bar & $E^\text{MD}$, eV & $\delta_E$,\% & $P^\text{MD}$, bar & $\delta_P$,\% \\ \hline
		14  & 4.40 & 0.72 & -2.881 & 3392.0 & -6.2(5) & 116 & 3034(28) & 11 \\
		20  & 1.51 & 0.51 & -0.353 & 1286.79 & -2.5(6) & 605 & 1209(9)  &  6 \\
		\hline
		25  & 0.77 & 0.40 & 1.372  & 703.67  & -0.3(7) & 122 & 670(6)   & 4.8 \\
		30  & 0.45 & 0.34 & 2.743  & 427.27  & 1.5(8)  & 47  & 412(5)   & 3.6 \\
		40  & 0.19 & 0.25 & 4.706  & 193.4  & 3.9(8)  & 18  & 187(3)   &  3.3 \\
		50  & 0.10 & 0.20 & 5.866  & 103.092  & 5.7(6)  & 2   & 102.1(9) &  1.0 \\
		60  & 0.06 & 0.17 & 6.563  & 61.0953   & 6.4(4)  & 2   & 60.5(3)  &  1.0 \\
		100 & 0.01 & 0.10 & 7.604  & 13.6866   & 7.4(2)  & 3   & 13.55(7) &  1.1 \\
		\hline
	\end{tabular}
\end{minipage}

\begin{minipage}{\textwidth}
\centering
\caption{Energy and pressure values for the isotherm $T=15625$~K (see caption to Table~\ref{tab:62500H}).}
\label{tab:15625H}
\begin{tabular}{|l|ll|ll|ll|ll|}
	\hline
	$r_s$ & $\chi \times 10^2$ & $\Gamma$ & $E^\text{PIMC}$, eV & $P^\text{PIMC}$, bar & $E^\text{MD}$, eV & $\delta_E$,\% & $P^\text{MD}$, bar & $\delta_P$,\% \\ \hline
	20  & 4.27 & 1.01 & -10.97 & 468.5 & -21.2(3) & 93 & 263(9) & 44 \\
	\hline
	40  & 0.53 & 0.51 & -10.06 & 61.95 & -20.0(4) & 98 & 36(2) & 42 \\
	60  & 0.16 & 0.34 & -8.99  & 19.336 & -16.2(9) & 80 & 13.1(5) & 32 \\
	80  & 0.07 & 0.25 & -7.90  & 8.611 & -12.3(4) & 56 & 6.8(3)  & 21 \\
	100 & 0.03 & 0.20 & -6.76  & 4.6378 & -11.3(5) & 68 &  4.0(2)  & 14\\
	\hline
\end{tabular}
\end{minipage}

\end{table*}

Before presenting our results for the non-degenerate hydrogen plasma at $\chi = 10^{-2}$, we validate our simulation method against the reference data obtained from first-principles PIMC simulations~\cite{Filinov:PhysRevE:2023, Bonitz:PP:2024}. 
Ref.~\onlinecite{Filinov:PhysRevE:2023} reports comprehensive tables containing the energy and pressure of hydrogen plasma along several isotherms as functions of the Brueckner parameter $r_s$ for 68 and 128 particles. Several results from these datasets relevant to our study are reproduced in Tables~\ref{tab:62500H}--\ref{tab:15625H}, corresponding to temperatures of $62.5$\,kK, $50$\,kK, $31.25$\,kK, and $15.625$\,kK.

In these tables, the first column reports the value of $r_s$, followed by the columns presenting the degeneracy parameter $\chi$ (multiplied by $100$ for convenience) and the coupling parameter $\Gamma$. The subsequent columns show the reference values of the internal energy (in eV) and pressure (in kbar) as computed in Ref.~\onlinecite{Filinov:PhysRevE:2023}. For direct comparison, the next two columns provide our MD results for the energy, and the respective differences with the reference. Similar structure is used for the pressure data. In our MD simulations, the system contained 68 particles (34 electrons) as in PIMC simulations of Ref.~\onlinecite{Filinov:PhysRevE:2023}.

Semiclassical MD simulations are limited by small values of the degeneracy parameter, $\chi \ll 1$. Accordingly, for values $10^2\times \chi < 1$, one may expect close agreement between the MD and PIMC simulation results, as the influence of quantum degeneracy is negligible. This is clearly observed in Table~\ref{tab:62500H}, where the deviations between MD and PIMC~\cite{Filinov:PhysRevE:2023} calculations remain below 1\% for both energy and pressure at temperatures $T~>~T_{H_2}$ and density parameter $r_s\geq 20$. However, with increasing degeneracy (decreasing $r_s$), these discrepancies become pronounced: at $r_s = 10$, the relative difference in energy approaches 10\%.

An additional example can be seen at 50~kK (see Table~\ref{tab:50000H}). For $r_s\geq 25$ ($10^2\times\chi < 1$), the differences between MD and PIMC results are larger. At $r_s=25$, the energy error is 3\%, while the pressure difference does not exceed 1\%. On the other hand, for the isotherm 31250~K (Table~\ref{tab:31250H}), the differences in the low degeneracy range become significant. For example, at $r_s=30$, the energy difference is 47\%, and at $r_s=25$, it reaches 100\%. In this case, the energy from MD is significantly lower than the value from the PIMC approach~\cite{Filinov:PhysRevE:2023}. 

\begin{figure}[h!]
	\centering
	\includegraphics[width=1\linewidth]{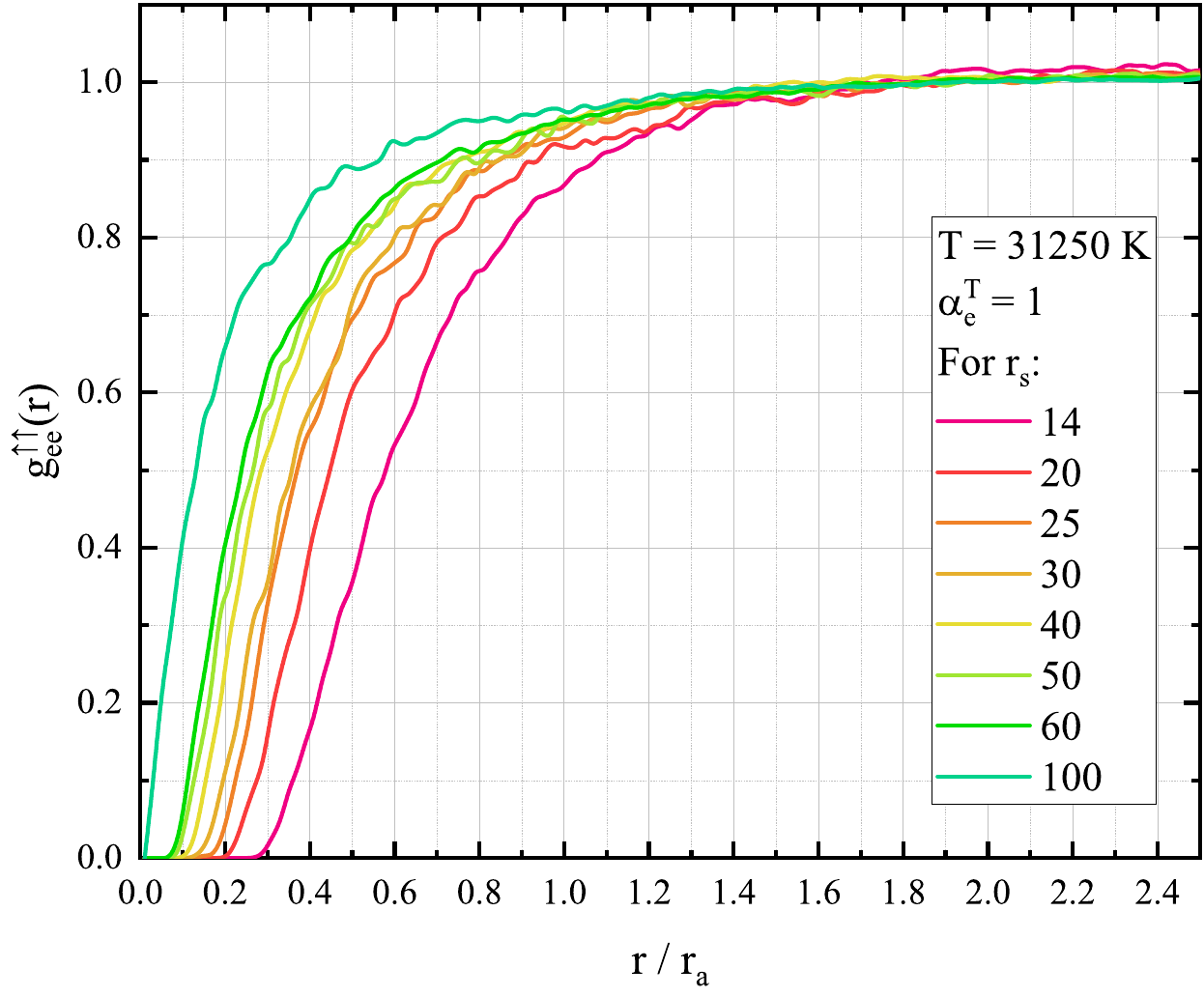}
	\caption{RDF for electrons with the same spin projection at  $T~=~31250$~K, obtained from MD simulations. The Pauli principle is obeyed, as no characteristic bound state peak appears (see Fig.~\ref{fig:spineebad}).}
	\label{fig:rdfeesamet31250}
\end{figure}

It is important to emphasize that such a considerable deviation \textit{is not caused by the appearance of unphysical cluster formation} arising from the bounding of electrons with identical spin projection, which was discussed in Ref.~\onlinecite{Filinov:PhysRevE:2004}. This is confirmed by the analysis of the electron RDF presented in Fig.~\ref{fig:rdfeesamet31250}. The peak that would indicate clusters as illustrated in Fig.~\ref{fig:spineebad} is not present in Fig.~\ref{fig:rdfeesamet31250}. Upon closely examining the plasma composition, as discussed below, we aim to explain this behavior.

Next, we examine the isotherm $T~=~15625$~K  (see Table~\ref{tab:15625H}). Across the entire parameter range studied, the energy is significantly underestimated by tens of percent compared to exact results. Despite the significant pressure differences, they remain several times smaller than the energy ones.

\begin{figure*}[ht!]
	\centering
	\begin{subfigure}[t]{0.49\linewidth}
		\includegraphics[width=\linewidth]{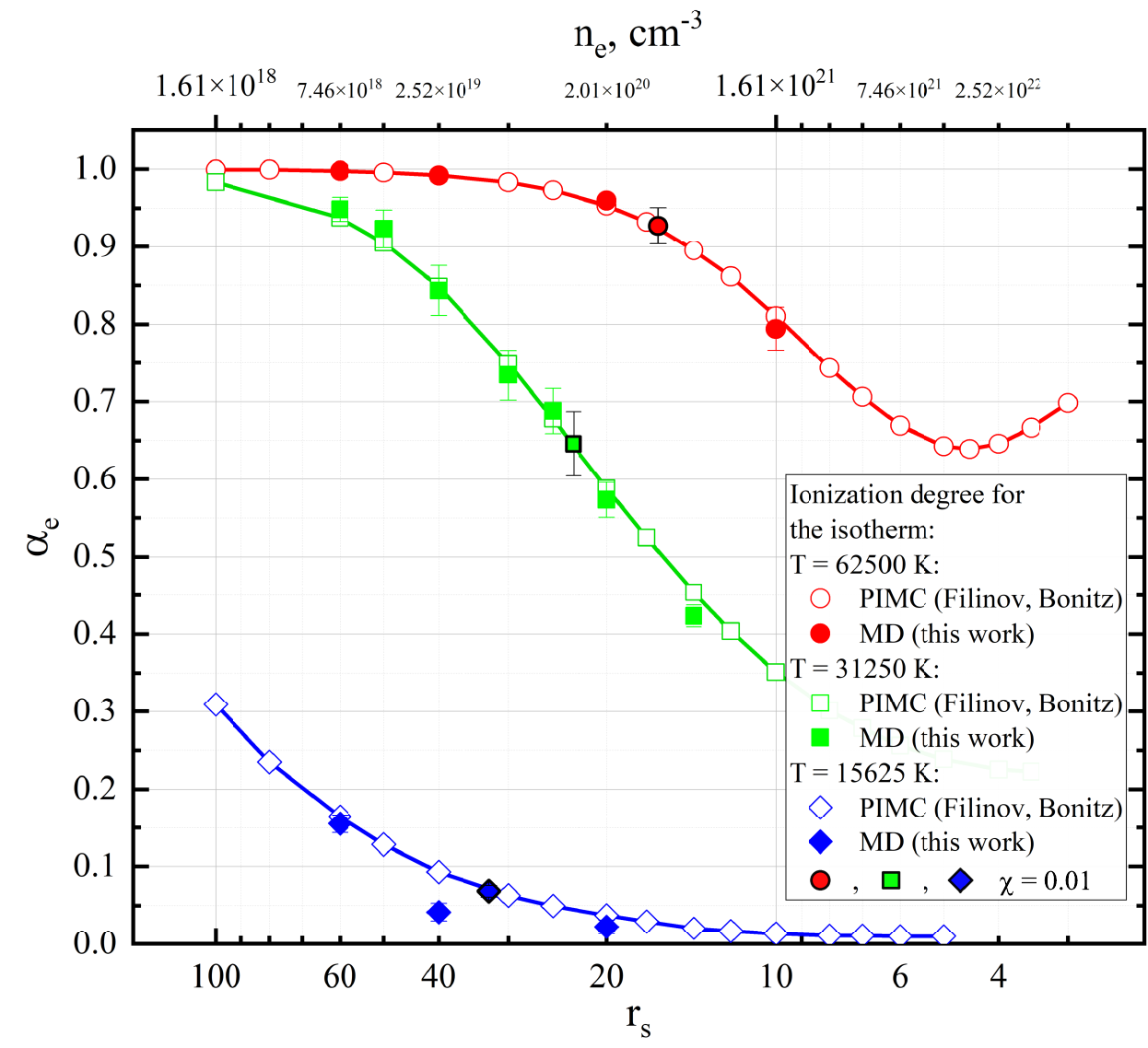}
		\caption{The ionization degree as a function of the parameter $r_s$ for three isotherms: $T~=~62500$~K, $T~=~31250$~K, and $T~=~15625$~K. Empty symbols show the data from Fig.~23~of~\onlinecite{Bonitz:PP:2024}, which was obtained based on the simulation data of~\onlinecite{Filinov:PhysRevE:2023}. Filled symbols show the MD data obtained in this work.}
		\label{fig:ion_coeff_filinov}
	\end{subfigure}
	\hfill
	\begin{subfigure}[t]{0.49\linewidth}
		\includegraphics[width=\linewidth]{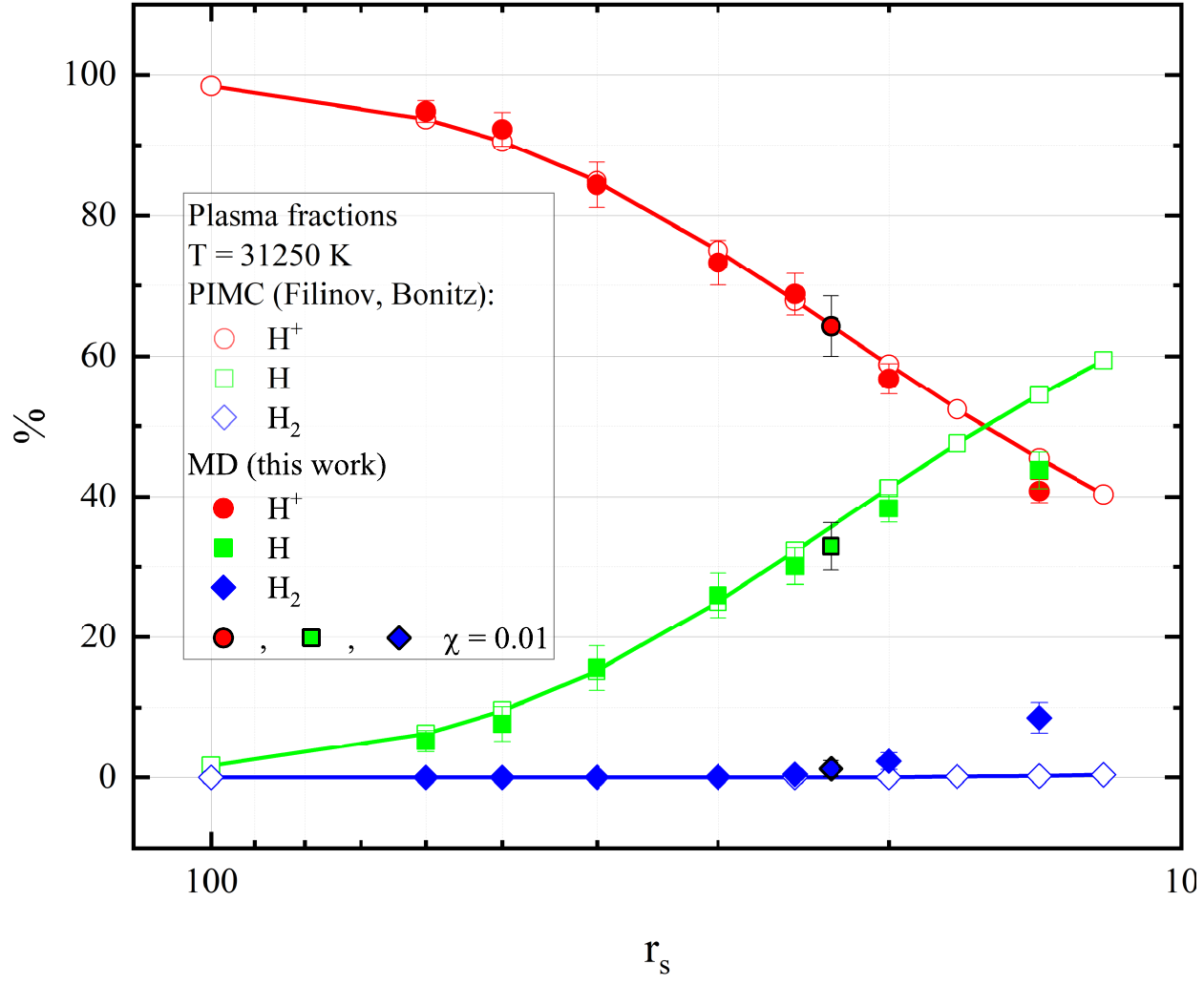}
		\caption{Fractions of free $H^+$ protons, $H$ atoms, and $H_2$ molecules at $T~=~31250$~K as a function of $r_s$. Empty symbols show data from Fig.~17~of~\onlinecite{Filinov:PhysRevE:2023}. Filled symbols denote MD data obtained in this work. }
		\label{fig:31250K_components}
	\end{subfigure}
	\caption{The ionization degree  of hydrogen plasma and its fraction for various isotherms depending on $r_s$. Empty symbols are the PIMC simulation data from Ref.~\onlinecite{Filinov:PhysRevE:2023}, and filled symbols are the MD simulation data of this work. $N = 68$ particles were used in the simulation. Filled symbols with a black border show the value at $\chi = 0.01$; symbols to the left of this correspond to $\chi < 0.01$, while those to the right indicate $\chi~>~0.01$.}
	\label{Fig:ioncoefffilinov}
\end{figure*}

We now turn to the analysis of the hydrogen plasma composition obtained from MD simulations and cluster analysis, including the ionization degree. The results for the ionization degree across several isotherms are presented in Fig.~\ref{fig:ion_coeff_filinov}, while Fig.~\ref{fig:31250K_components} shows the plasma composition at $T~=~31250$~K. For comparison, the PIMC data are taken from Refs.~\onlinecite{Filinov:PhysRevE:2023, Bonitz:PP:2024}.

Let us examine the behavior of the ionization degree in Fig.~\ref{fig:ion_coeff_filinov}. It is evident that the values of $\alpha_e$ agree well for $r_s~>~20$. This range corresponds to the weak degeneracy regime for both isotherms $T~=~62500$~K and $T~=~31250$~K. This agreement between MD and PIMC shows that the criterion for selecting the threshold $d_H$ to identify electrons in bound states (see Section~\ref{sec:algcomplex}) is correct in the non-degenerate plasma regime.

However, the ionization degree is twice lower compared to the PIMC calculations for $T~=~15625$~K. This means that an excessive number of electrons are in bound states compared to what is necessary. Such behavior indicates that the attractive forces between electrons and protons are too strong compared to the exact ones. 

Next, we consider the plasma composition dependence on $r_s$ along the $T=31250$~K isotherm. Figure~\ref{fig:31250K_components} presents the fraction of free protons, atoms, and molecules as functions of $r_s$, obtained from the MD simulation and compared with the data from Ref.~\onlinecite{Filinov:PhysRevE:2023}. It is clear that in the region $\chi < 0.01$ (or equivalently $r_s \geq 25$) the fractions from MD and PIMC agree well.

However, for $r_s \leq 30$, as $r_s$ decreases, molecules appear in MD simulations. These molecules are absent in the PIMC data. Simultaneously, the fraction of atoms decreases significantly. This behavior indicates that a large number of atoms are bonding into molecular compounds. The result shows again that the attraction between protons and electrons is too strong.

Note that our MD simulations also exhibit the formation of $H_2^+$ complexes besides molecules. For clarity, these are not shown in Fig.~\ref{fig:31250K_components}.

The observation that MD simulations employing the improved Kelbg p/p result in an excessive drop of energy was already reported twenty years ago~\cite{Filinov:PhysRevE:2004}. In those calculations, the Pauli principle was violated. This violation caused the unphysical formation of bound states of electrons with the same spin projection.

However, in Ref.~\onlinecite{Filinov:PhysRevE:2004}, a moderately degenerate plasma was considered. Thus, this behavior was attributed to many-particle quantum effects not considered in the improved Kelbg p/p and additional spin terms.

In this paper, we consider a non-degenerate system with $\chi \ll 1$. In this regime, many-particle \textit{quantum} effects are not important. Inside a molecule, i.e., at small distances, there are only two electrons with the opposite spin projections. Meanwhile, the electrons with the same spin projection are significantly more distanced than $\lambda_{e}$.

To prevent electrons with the same spin projection from approaching each other in MD simulations, we propose taking into account the finite size of electrons (see Eq.~\eqref{eq:fdrgdfsef}). This modification effectively excludes the unphysical approach of electrons with identical spin projections in non-degenerate plasma. As a result, in all calculations presented in this section, we do not observe the formation of nonphysical complexes discussed in Ref.~\onlinecite{Filinov:PhysRevE:2004}.

However, the total energy of the system is found to be significantly lower than the correct value. Furthermore, increasing the density leads to the excessive formation of molecular compounds compared to quasi-exact PIMC simulations~\cite{Filinov:PhysRevE:2023, Bonitz:PP:2024}.

Since there are no bound states formed by electrons with the same spin projection, the issue is likely due to the improper behavior of the improved Kelbg p/p during molecular formation. The study in Ref.~\onlinecite{Filinov:PhysRevE:2004} showed that this p/p reproduces the exact density matrix behavior with excellent  accuracy even at relatively low temperatures down to 5~kK (see Fig.~5 in Ref.~\onlinecite{Filinov:PhysRevE:2004}). 
Nevertheless, subtle differences in the p/p that are not captured by the expression of the improved Kelbg p/p may significantly affect the particle configurations formed by interparticle forces. 

In other words, if we denote the exact p/p derived from the density matrix as $\Phi^\text{ex}_0(r;\beta)$, then the values match approximately: $\Phi^\text{ex}_0(r;\beta) \approx \Phi^\text{I}_0(r;\beta)$. However, their gradients may differ significantly, so that $\nabla \Phi^\text{I}_0(r;\beta)$ provides a poor approximation to the function $\nabla \Phi^\text{ex}_0(r;\beta)$.

This paper does not prove this hypothesis, since calculating the exact Coulomb density matrix at arbitrary temperature is beyond its scope. Note that the need to increase the parameter $\alpha_e^T$ for electrons with increasing coupling parameter $\Gamma$ (see Tab.~\ref{tab:alphaGamma}) indirectly shows inaccuracies in the improved Kelbg p/p. The investigation of this hypothesis is the subject of our future work.

Despite these limitations, we will present the results obtained using the improved Kelbg p/p that accounts for the long-range interaction dependent on the coupling parameter $\Gamma$, fixing the degeneracy parameter at $\chi~=~0.01$. Although the accuracy of such simulation in the low-temperature regime is low, as demonstrated above, it still captures the qualitative behavior of the system under these conditions. It also reveals possible issues that might arise even with more accurate simulations.

\subsection{Radial distribution functions and plasma composition for $\chi = 0.01$}

\begin{figure*}[ht!]
	\centering
	\begin{subfigure}[t]{0.49\linewidth}
		\includegraphics[width=\linewidth]{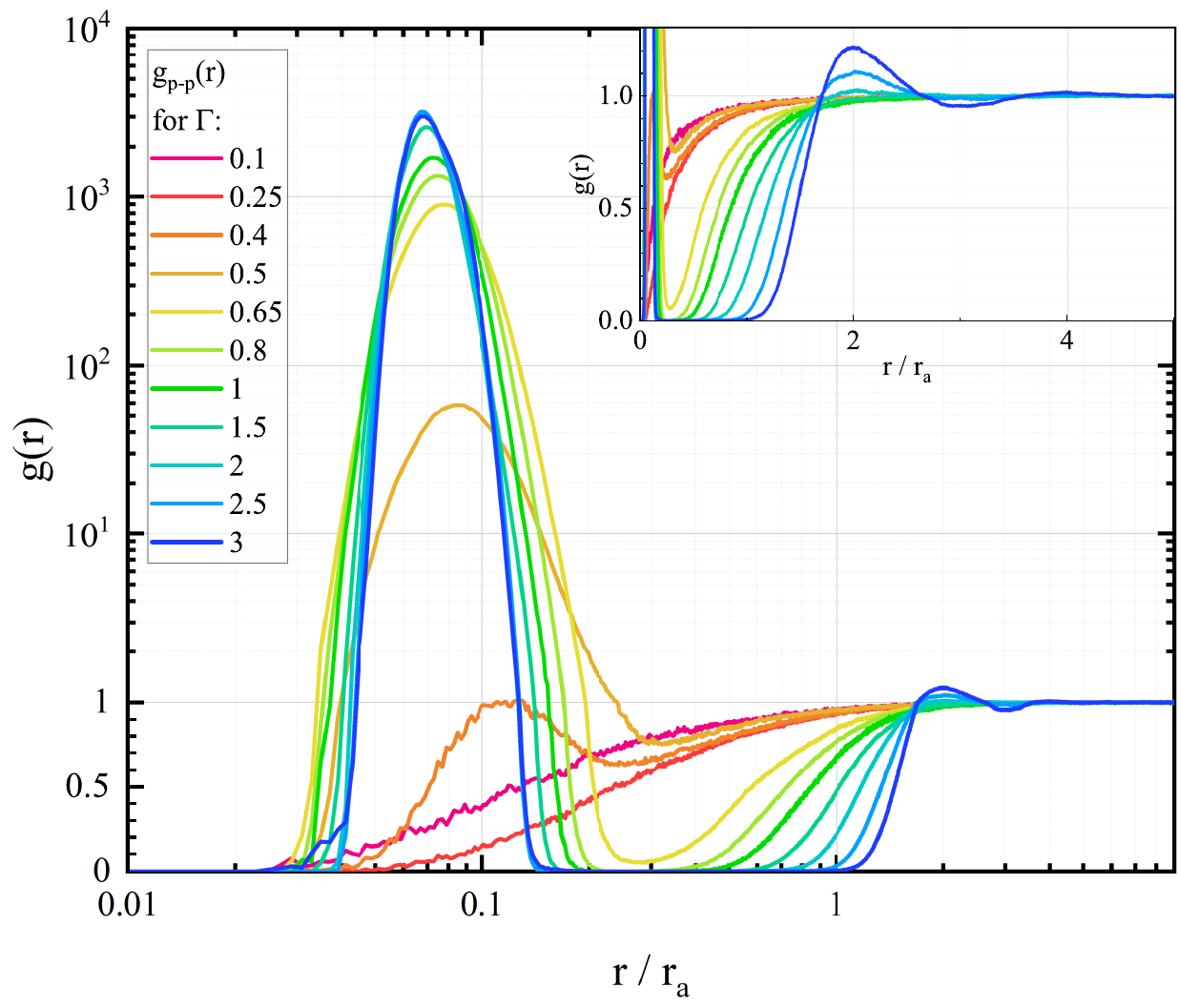}
		\caption{The RDF $g_{p-p}(r)$ for all protons in the system.}
		\label{fig:rdf_pp}
	\end{subfigure}
	\hfill
	\begin{subfigure}[t]{0.49\linewidth}
		\includegraphics[width=\linewidth]{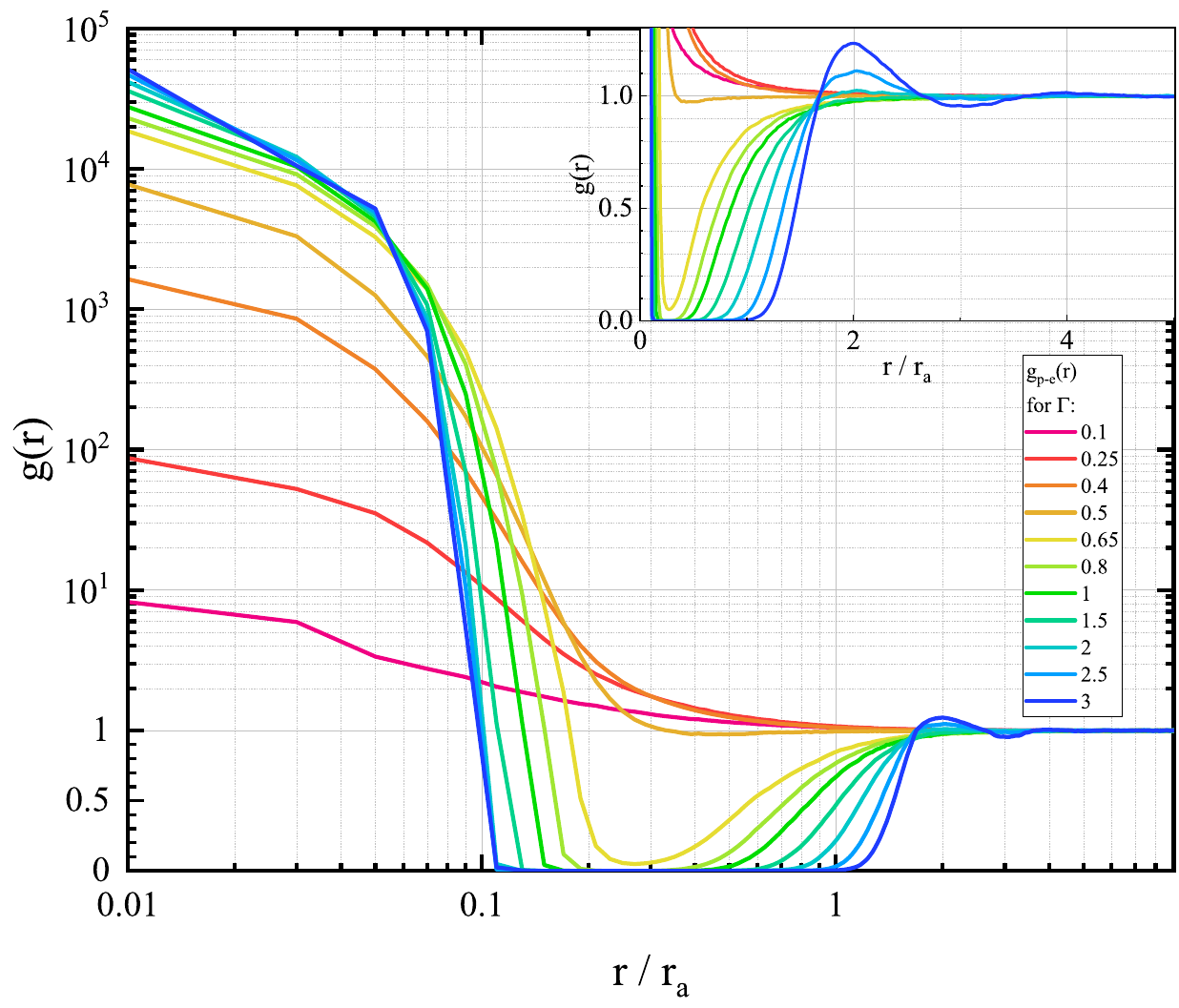}
		\caption{The RDF $g_{p-e}(r)$ for all protons and electrons.}
		\label{fig:rdf_ep}
	\end{subfigure}
	
	\begin{subfigure}[b]{0.49\linewidth}
		\includegraphics[width=\linewidth]{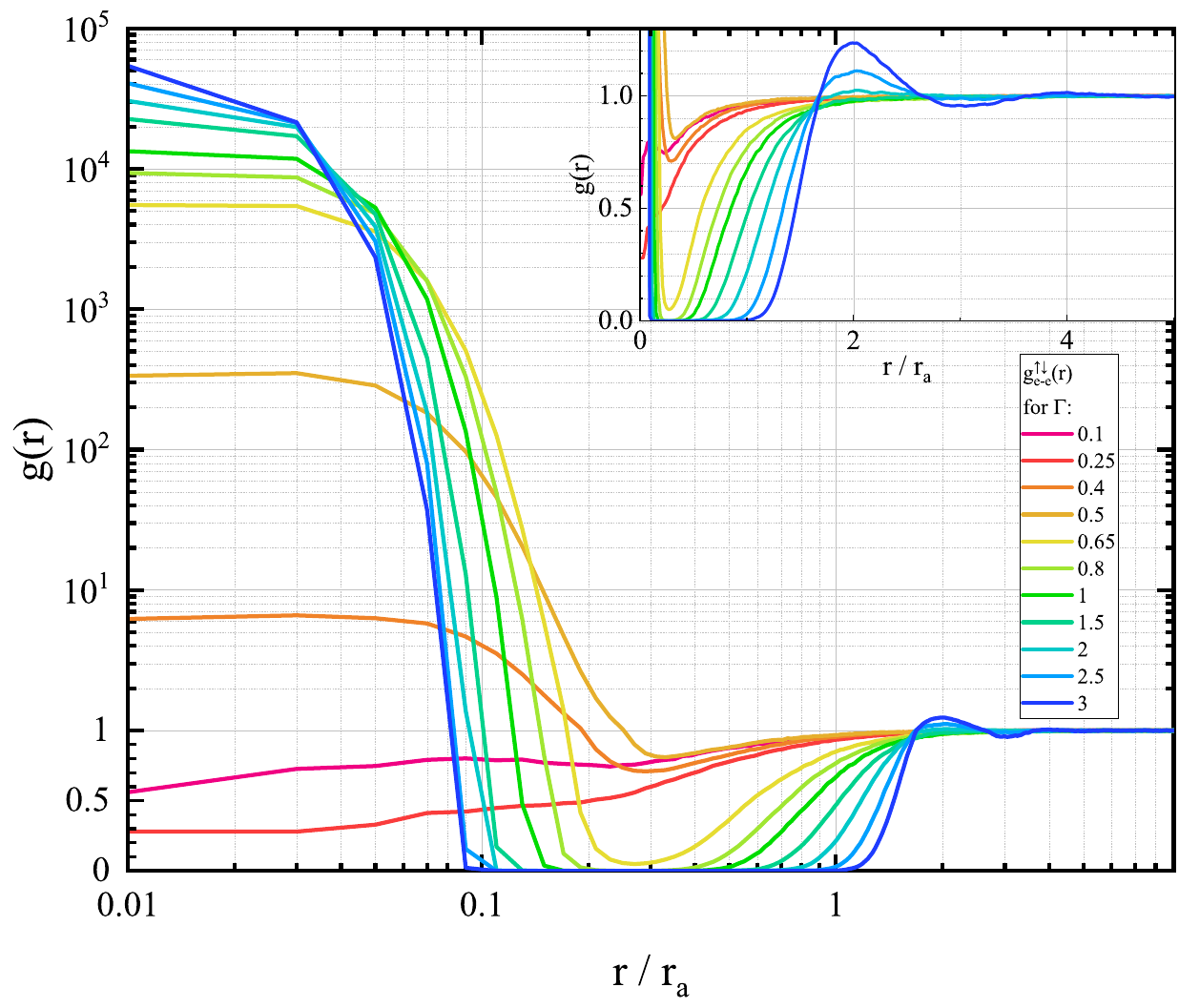}
		\caption{The RDF $g_{e-e}^{\uparrow\downarrow}(r)$ for electrons with the opposite spin projections.}
		\label{fig:rdf_ee_diff}
	\end{subfigure}
	\hfill
	\begin{subfigure}[b]{0.49\linewidth}
		\includegraphics[width=\linewidth]{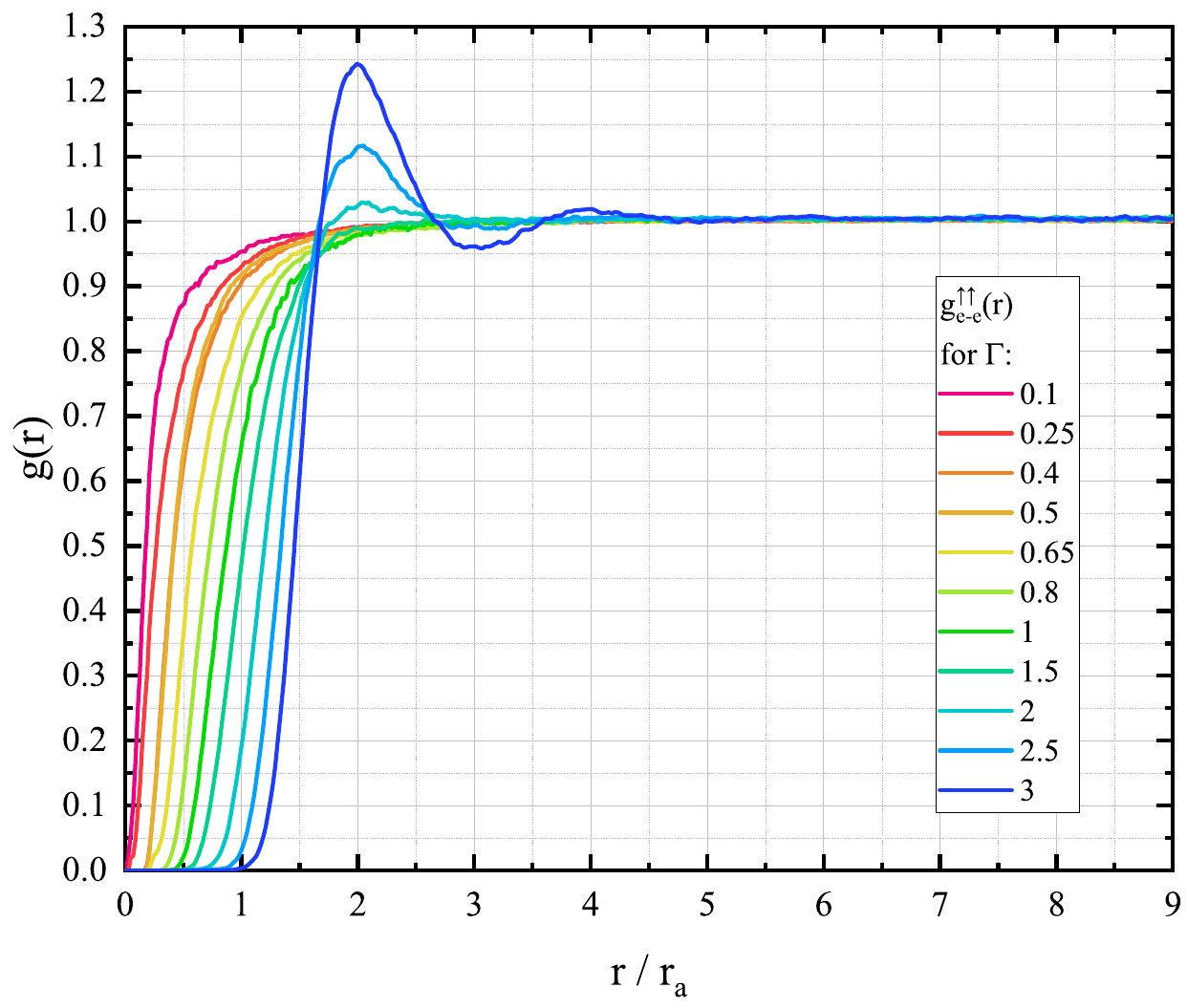}
		\caption{The RDF $g_{e-e}^{\uparrow\uparrow}(r)$ for electrons with the same spin projection.}
		\label{fig:rdf_ee_same}
	\end{subfigure}
	\caption{Radial distribution functions for non-degenerate hydrogen plasma at $\chi = 0.01$, obtained using MD simulations. In panels (a)--(c), the horizontal axis is logarithmic. The vertical axis is linear from 0 to 1, and logarithmic for $g(r) \geq 1$. In the upper right corner of panels (a)--(c), linear-scale insets are provided for both axes. Simulations were performed with $N = 10^3$ particles.
	}
	\label{Fig:rdf_h}
\end{figure*}

\begin{figure*}[ht!]
	\centering
	\begin{subfigure}[t]{0.49\linewidth}
		\includegraphics[width=\linewidth]{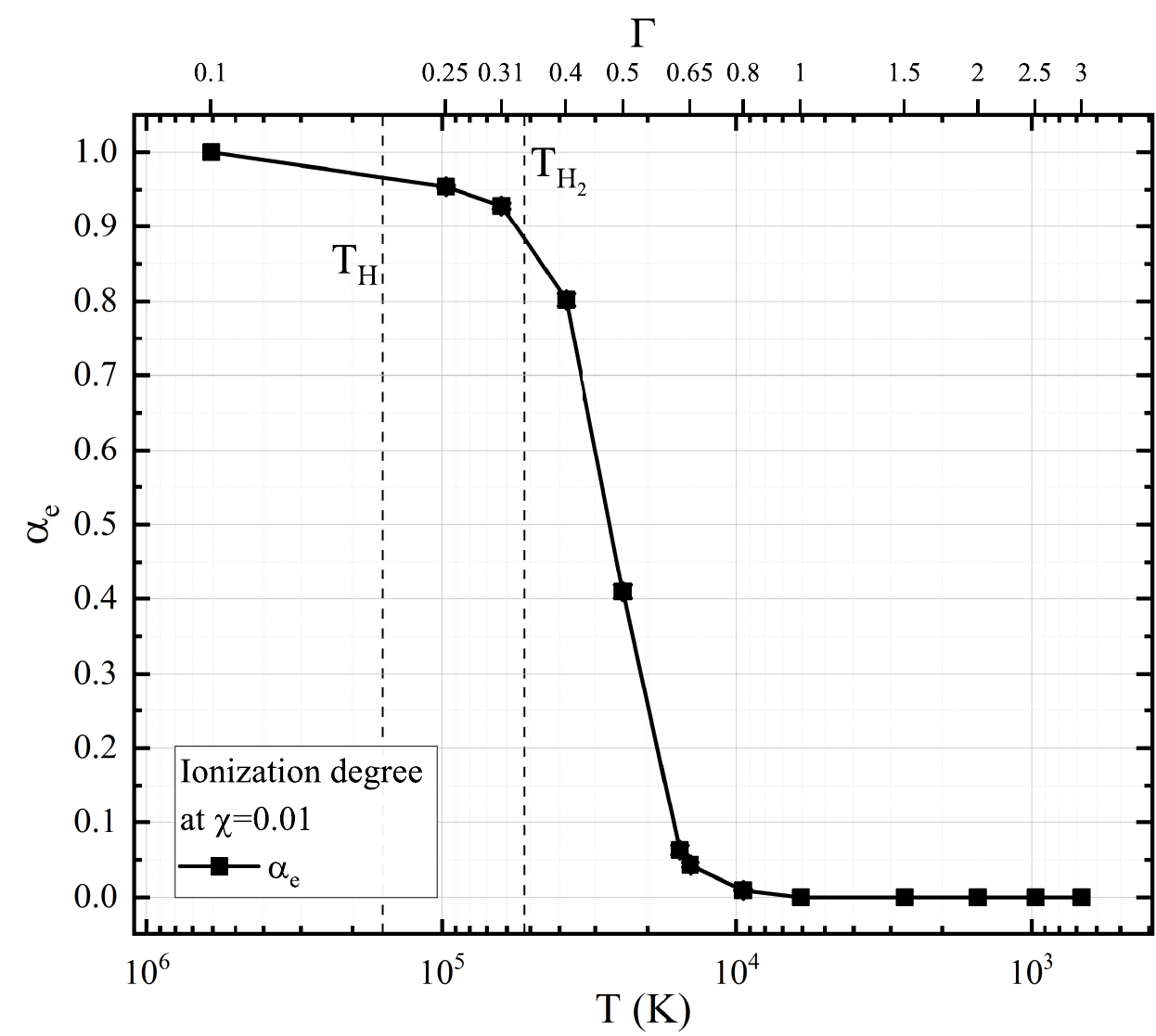}
		\caption{The ionization degree of non-degenerate hydrogen plasma as a function of temperature and the coupling parameter.}
		\label{fig:ion_coeff_chi}
	\end{subfigure}
	\hfill
	\begin{subfigure}[t]{0.49\linewidth}
		\includegraphics[width=\linewidth]{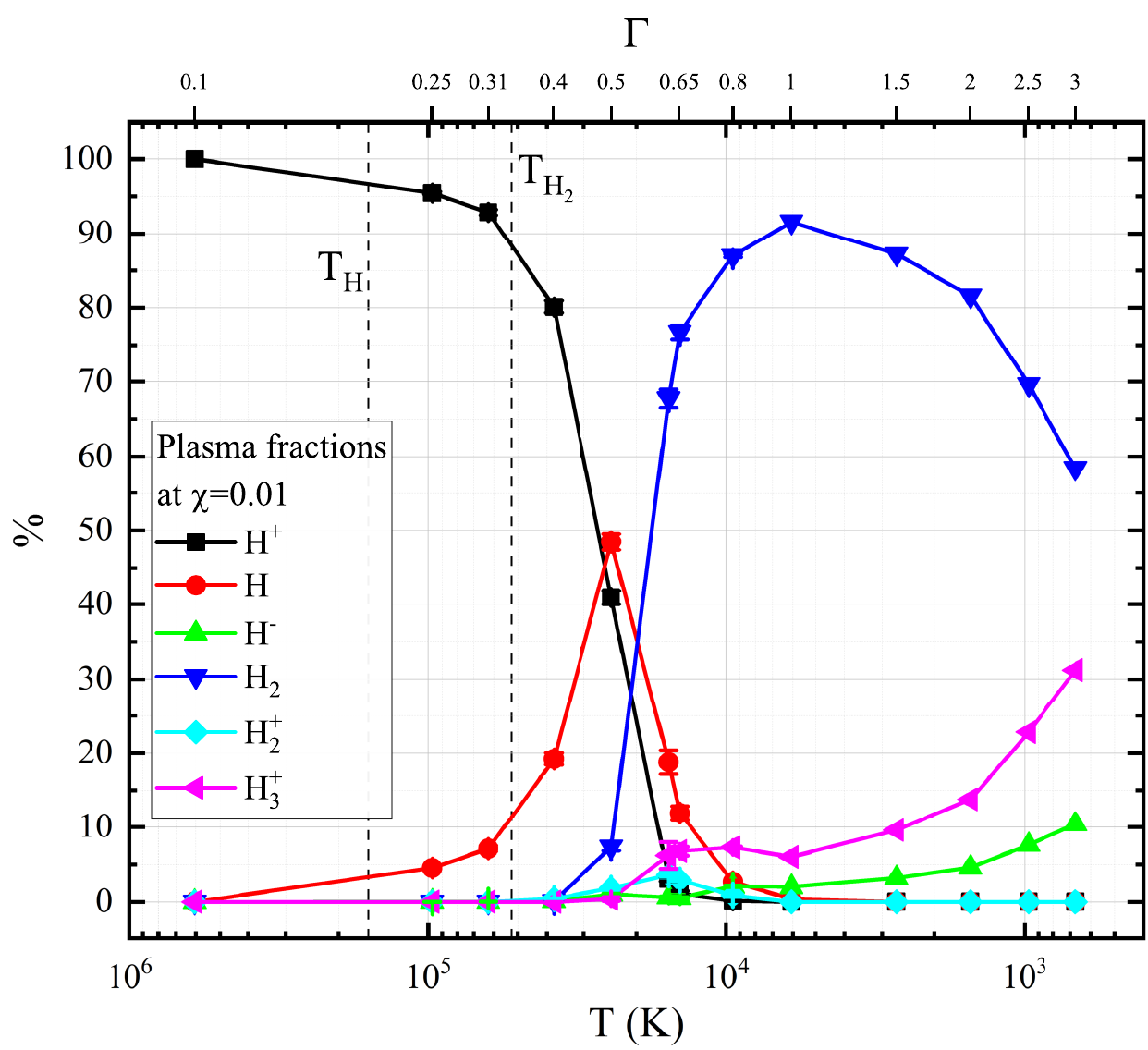}
		\caption{The composition of non-degenerate hydrogen plasma as a function of temperature and the coupling parameter.}
		\label{fig:components_chi}
	\end{subfigure}
	\caption{Ionization degree and fraction of non-degenerate hydrogen plasma at $\chi = 0.01$. Calculations are shown for $N = 10^3$.}
	\label{Fig:ioncoeffchi}
\end{figure*}

To investigate the structure of hydrogen plasma as well as to analyze the plasma composition, we examine its RDFs. These include the RDF for protons, $g_{p-p}(r)$, the RDF between protons and electrons, $g_{p-e}(r)$, the RDF between electrons with the opposite spin projections, $g_{e-e}^{\uparrow\downarrow}(r)$, and the RDF for electrons with the same spin projection, $g_{e-e}^{\uparrow\uparrow}(r)$.

The $g_{p-p}(r)$ plot (Fig.~\ref{fig:rdf_pp}) shows that there are no molecules present  for $\Gamma \leq 0.25$. This can be explained by the fact that the temperature ($T \ge 97$~kK, see Table~\ref{tab:NdepH}) is above the dissociation energy $T_{H_2}$. At $\Gamma = 0.5$, a bound state peak appears, indicating the onset of bound states between two protons and the formation of molecular compounds. 

For $\Gamma \leq 0.5$, the RDF~\ref{fig:rdf_pp} remains greater than zero at $r \geq 0.1$. This behavior changes significantly for $\Gamma \geq 0.8$. After the bound state peak, which is evident at $r \approx 0.07 r_a$, the function $g_{p-p}(r)$ drops sharply to zero and then approaches unity only for $r~>~r_a$. This behavior results from the increase in coupling. Therefore the system is less dense (see Table~\ref{tab:NdepH}) and molecular compounds rarely collide with each other.

At $\Gamma \geq 2$, the RDF for protons~\ref{fig:rdf_pp} shows both maximum and minimum at $r > r_a$. This behavior indicates the appearance of short-range order in the system.

The RDFs for electrons and protons (Fig.~\ref{fig:rdf_ep}), as well as for electrons with the opposite spin projections (Fig.~\ref{fig:rdf_ee_diff}), exhibit similar behavior at large distances. As the coupling parameter $\Gamma$ increases, a region appears where  $g(r)$ is equal to zero. For $\Gamma \geq 2$, the first maximum appears at $r>r_a$.

In contrast, electrons with the same spin projection do not form a bound state. As a result, $g_{e-e}^{\uparrow\uparrow}(r)$ rapidly approaches 1, even for $\Gamma \leq 1.5$. For larger values of $\Gamma$, specifically for $\Gamma \geq 2$, the RDF demonstrates pronounced maximum and minimum values. These indicate the onset of short-range order in the system.

Following the procedure described in Sec.~\ref{sec:algcomplex}, we calculated the plasma composition and ionization degree $\alpha_e$ for non-degenerate hydrogen as functions of the coupling parameter~$\Gamma$. Figure~\ref{fig:ion_coeff_chi} shows that $\alpha_e$ is equal to one for $T~>~T_H$, but decreases as the temperature drops below $T_H$. When the temperature is further reduced to $T_{H_2}$, the ionization degree drops more rapidly. It reaches the value of $0.5$ at $30$~kK (corresponding to $\Gamma = 0.4$). At $\Gamma > 0.8$, $\alpha_e$ vanishes completely.

Figure~\ref{fig:components_chi} presents the composition of a non-degenerate hydrogen plasma. The figure shows the fraction of plasma components as a function of temperature and the coupling parameter $\Gamma$. It is evident that the fraction of free protons closely follows the ionization degree. As the temperature decreases in the region $T_H < T < T_{H_2}$, free electrons and protons are bound only into atoms. Molecules and other complexes are absent in this region.

As the temperature decreases from $40$ to $25$~kK, the formation of molecules can be observed. There are also a small number of ionized complexes, such as $H^-$ and $H_2^+$. As temperature further decreases from $25$ to $10$~kK, the atom fraction decreases rapidly. At the same time, the number of molecules increases significantly and reaches a maximum near $6$~kK ($\Gamma = 1$). When the temperature drops below this value, the number of molecules starts to decrease. The proportion of ionized complexes, such as $H_3^+$ and $H^-$, increases significantly.

Thus, for $\Gamma \geq 0.8$, free electrons and protons disappear completely, and the fraction of molecules and complexes exceeds the number of atoms. Bound states dominate the system in this regime.

Therefore, for $\Gamma \geq 0.8$, the simulation yields a nearly constant plasma composition that remains unchanged throughout the MD. This behavior causes some difficulties in reaching the thermodynamic limit at $\Gamma \geq 0.8$, as will be demonstrated in the next section. 

\subsection{$N$-dependence of energy and thermodynamic limit at $\chi = 0.01$}
\begin{figure*}[ht!]
	\centering
	\begin{subfigure}[b]{0.49\linewidth}
		\includegraphics[width=\linewidth]{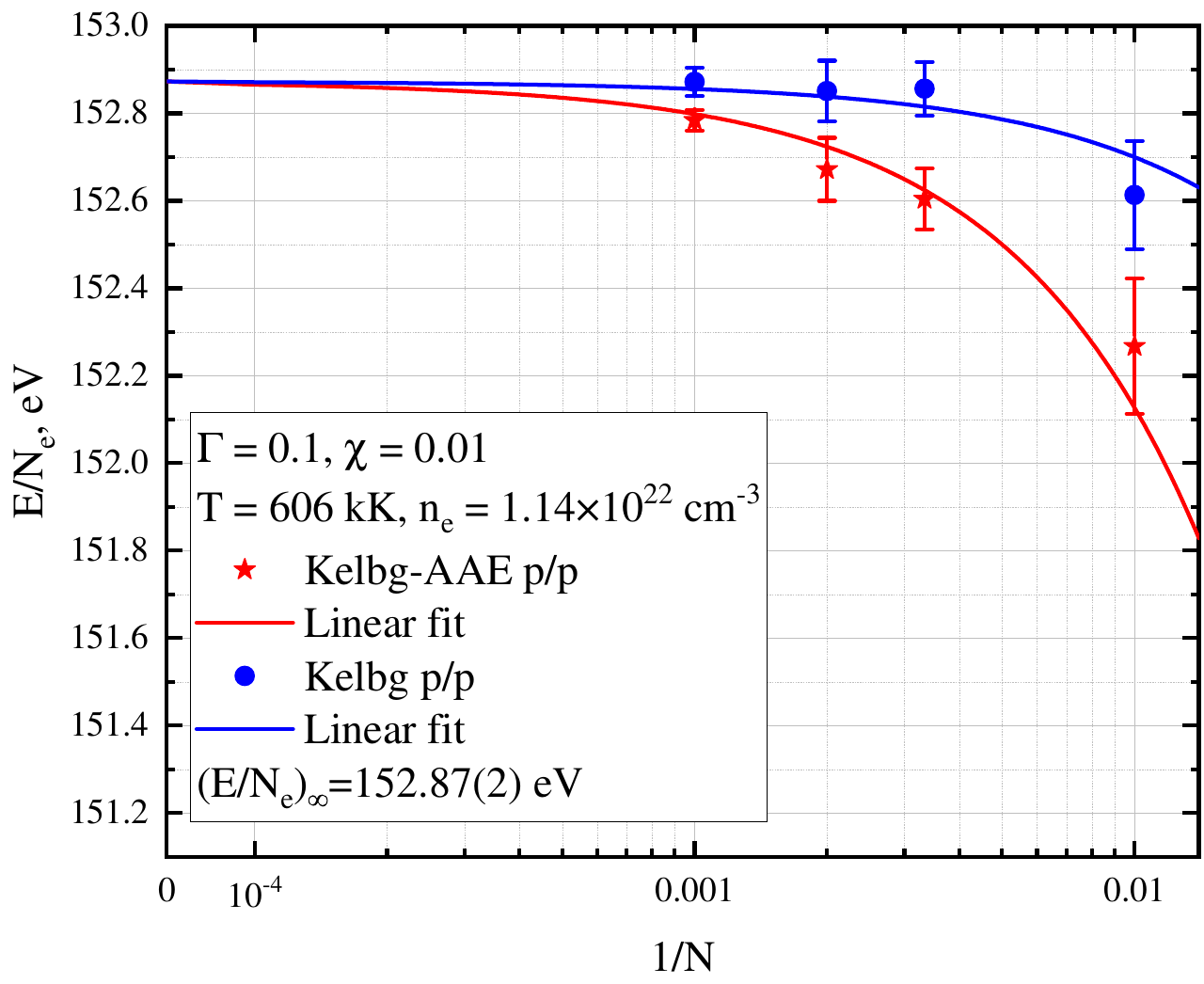}
		\caption{$\Gamma = 0.1$}
		\label{fig:Ndep1}
	\end{subfigure}
	\hfill
	\begin{subfigure}[b]{0.49\linewidth}
		\includegraphics[width=\linewidth]{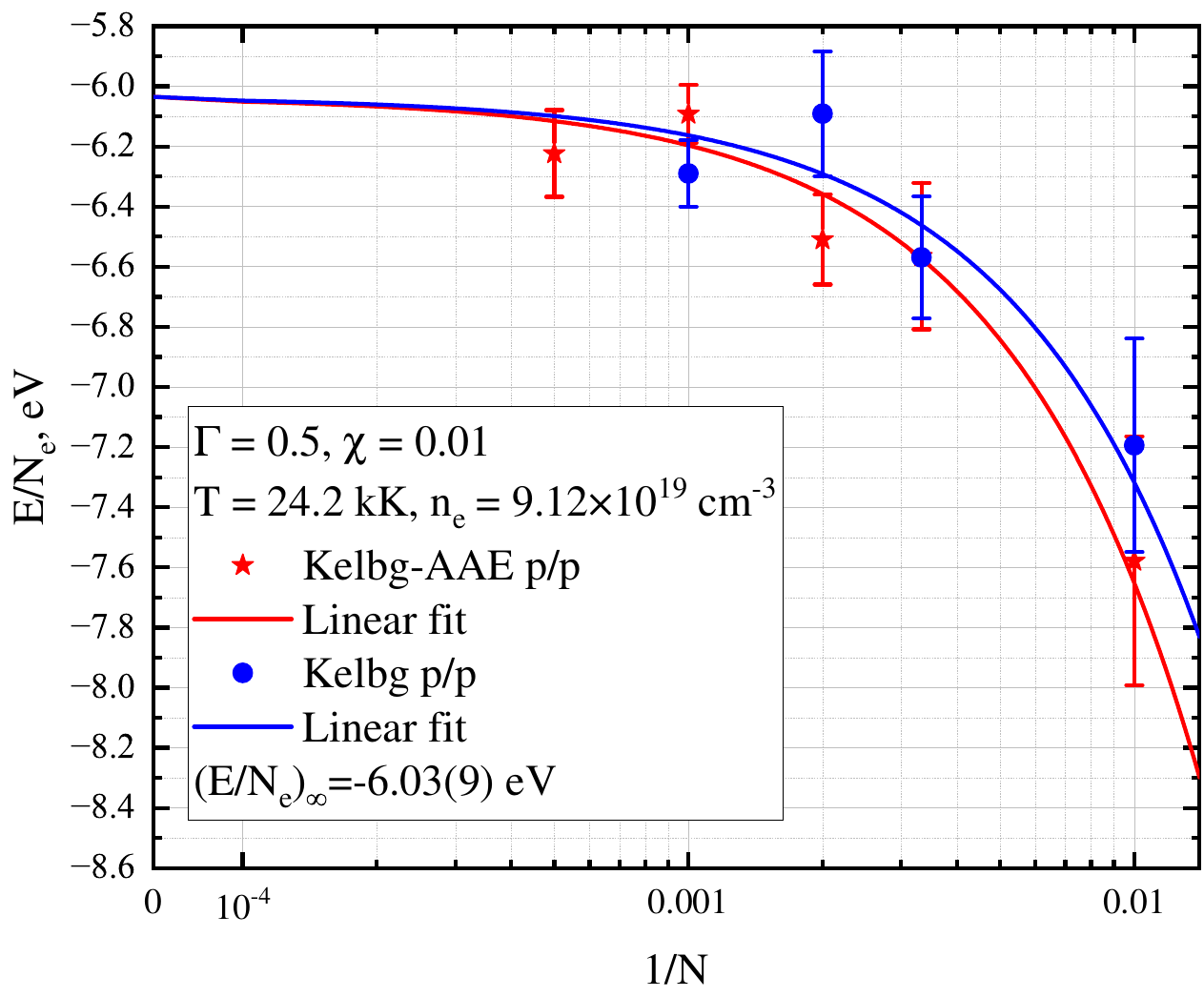}
		\caption{$\Gamma = 0.5$}
		\label{fig:Ndep4}
	\end{subfigure}
	
	\begin{subfigure}[b]{0.49\linewidth}
		\includegraphics[width=\linewidth]{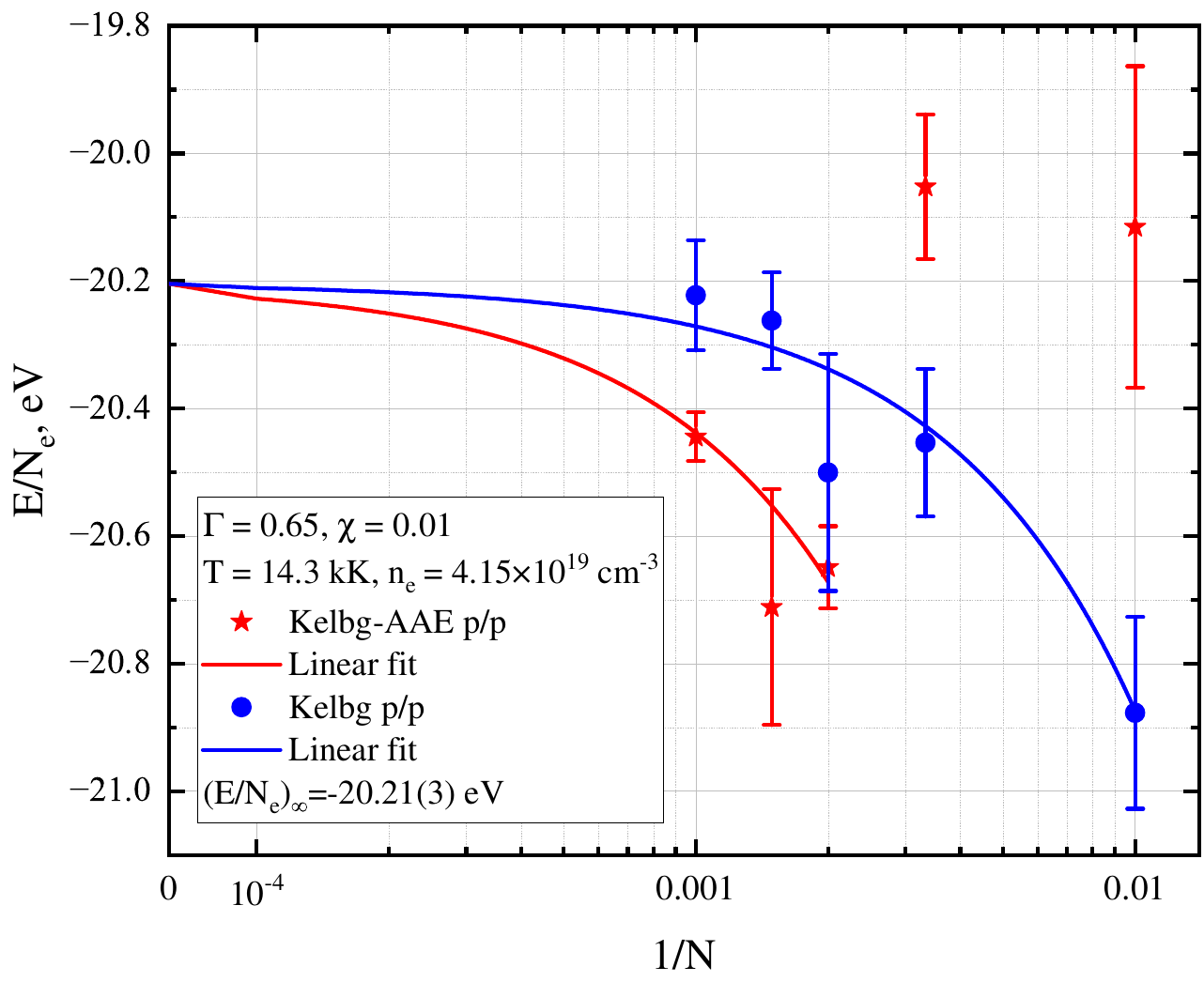}
		\caption{$\Gamma = 0.65$}
		\label{fig:Ndep1H}
	\end{subfigure}
	\hfill
	\begin{subfigure}[b]{0.49\linewidth}
		\includegraphics[width=\linewidth]{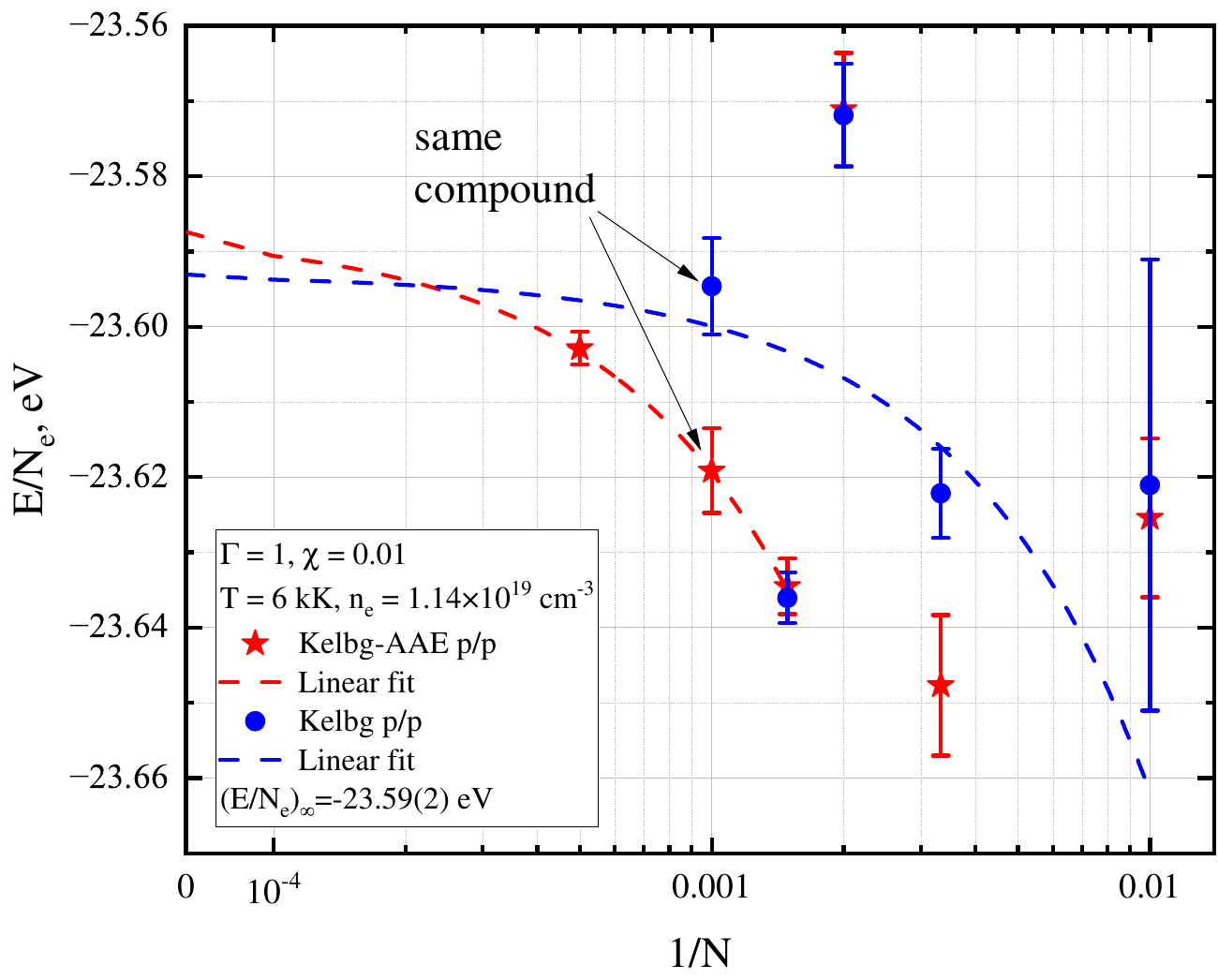}
		\caption{$\Gamma = 1$}
		\label{fig:Ndep2H}
	\end{subfigure}
	\caption{Dependence of the total energy per electron of a non-degenerate hydrogen plasma on $N^{-1}$ in the region $0.1\leq \Gamma \leq 1$ for $\chi = 0.01$. The logarithmic scale is used on the horizontal axis in the interval $(10^{-4}, 0.014)$; the linear scale is used for $1/N <10^{-4}$. One can see that for strong coupling the dependence on the number of particles is jumpy. Due to this behavior, it is not possible to obtain a reliable thermodynamic limit using $N\leq 10^3$ for $\Gamma \geq 0.65$.}
	\label{Fig:NdepG}
\end{figure*}

In this section, we analyze the calculation of the thermodynamic limit for the energy of a non-degenerate hydrogen plasma. We consider different values of the coupling parameter $\Gamma$ and examine the results obtained both with and without including long-range interactions. We also discuss specific aspects of evaluating the thermodynamic limit for $\Gamma \geq 0.8$.

Figure~\ref{Fig:NdepG} shows the energy per electron as a function of the number of particles. The calculations were performed using the improved Kelbg-AAE p/p and improved Kelbg p/p (see Eq.~\eqref{eq:diagKelbgAAEPPImroved}). 

First, we focus on the case $\Gamma \leq 0.5$, where particle collisions are frequent (see Figs.~\ref{fig:Ndep1} and~\ref{fig:Ndep4}). In this regime, the $N$-dependence of energy is smooth. Therefore, it can be accurately fitted using a linear approximation to obtain the limit $N^{-1}\to 0$. 

Figures~\ref{fig:Ndep1H}~and~\ref{fig:Ndep2H} demonstrate that the smooth dependence on the number of particles, which is observed for $\Gamma \leq 0.5$, disappears at larger $\Gamma$. The data points are scattered irregularly (or jumpy), especially for small values of $N$. There are at least two possible reasons for this behavior.

First, the simulation could achieve thermodynamic equilibrium but not chemical equilibrium. In this case, a specific composition forms randomly and then remains unchanged due to the high dilution of the system. One might propose that the simulation time is insufficient to reach chemical equilibrium. However, a typical simulation duration on the equilibrium section, as given by Eq.~\eqref{eq:dgdtbdtdrf}, is $\sqrt{200}\frac{\sqrt{3}}{2\pi}10^7\times 5\times 10^{-4}\tau_p = 19000\tau_p$. This duration corresponds to tens of thousands of plasma periods. To verify this, the simulation time was increased by an order of magnitude for some selected points. Nonetheless, the composition did not change even during this extended simulation. Therefore, it is reasonable to conclude that chemical equilibrium was achieved, including the points illustrated in Fig.~\ref{Fig:NdepG}.

Second, the chosen number of particles is non-optimal given the thermodynamic conditions. In small, fixed-size systems, achieving the optimal composition can be challenging due to the limited number of particles of each type. Consequently, the fractions of the components may either change abruptly as the number of particles, $N$, increases, or remain unchanged. This behavior contrasts with the smooth dependence on $N$ that is typically expected in systems with a larger number of particles (or in the absence of bound states as in Ref.~\onlinecite{Demyanov:CPP:2024}).

This issue can be addressed by either increasing the number of particles, for example, using $N \geq 10^3$, or by employing simulation methods that allow for a variable number of particles. However, MD simulations involving more than a thousand particles can become time-consuming, as the time required to achieve chemical equilibrium increases significantly. Additionally, using methods with a variable number of particles is beyond the scope of this study.

Although Fig.~\ref{fig:Ndep2H} does not exhibit a smooth dependence, note that the differences in energy between different $N$ are quite small. For example, they are about $0.1\%$ at $\Gamma \geq 1$. Therefore, one could estimate the thermodynamic limit using the energy corresponding to the largest $N$ in the simulations or by fitting several selected data points. In Fig.~\ref{fig:Ndep2H},  a dotted line is shown through some selected points, using the fits that are linear and quadratic in $1/N$.

Finally, for $\Gamma \geq 0.65$, the difference between the calculations with and without long-range interactions becomes negligible. In this regime, the main contribution to the total energy comes from the short-range part of the p/p. As a result, the computed energy values are almost the same, regardless of whether the improved Kelbg p/p is used with or without the long-range correction.

\begin{table}[!ht]
	\caption{Energy and pressure of non-degenerate hydrogen plasma at $\chi = 0.01$ in the thermodynamic limit. The uncertainty of the last digits is shown in brackets.}
	\label{tab:NdepH}
	\begin{tabular}{|c|c|c|c|c|c|}
		\hline
		$\Gamma$ & $T$, kK & $n_e$, cm$^{-3}$ & $(E/N_e)_{\infty}$, eV & $(P)_\infty$, bar & $r_s$ \\ \hline
		$0.1$ & $606$ & $1.14\times 10^{22}$ & $152.87(2)$ & $1885.7(4)\times 10^3$ & $5.21$ \\ 
		$0.25$ & $97.0$ & $7.29\times 10^{20}$ & $22.17(2)$ & $18.509(8)\times 10^3$ & $13.0$ \\ 
		$0.4$ & $37.9$ & $1.78\times 10^{20}$ & $4.1(2)$ & $1725(6)$ & $20.8$ \\ 
		$0.5$ & $24.2$ & $9.12\times 10^{19}$ & $-6.03(9)$ & $456(2)$ & $26.0$ \\ 
		$0.65$ & $14.3$ & $4.15\times 10^{19}$ & $-20.21(3)$ & $82.4(13)$ & $33.9$ \\ 
		$0.8$ & $9.47$ & $2.23\times 10^{19}$ & $-22.60(2)$ & $34.0(12)$ & $41.7$ \\ 
		$1$ & $6.06$ & $1.14\times 10^{19}$ & $-23.59(2)$ & $12.6(3)$ & $52.1$ \\ 
		$1.5$ & $2.69$ & $3.38\times 10^{18}$ & $-24.715(3)$ & $2.939(8)$ & $78.1$ \\
		$2$ & $1.52$ & $1.42\times 10^{18}$ & $-25.3068(3)$ & $1.18(2)$ & $104$ \\
		$2.5$ & $0.97$ & $7.29\times 10^{17}$ & $-25.623(5)$ & $0.632(9)$ & $130$ \\ \hline
	\end{tabular}
\end{table}

As a result, we conclude that taking into account long-range interaction does not improve the convergence of energy on the number of particles. For small $\Gamma$, as in the case of the one-component plasma~\cite{Demyanov:CPP:2024}, a faster $N$-convergence is provided by a p/p without long-range interaction, and for large $\Gamma$ the energy is mainly determined by the behavior at small distances, which is the same for both pseudopotentials. We have collected the obtained thermodynamic limit for energy and pressure in Tab.~\ref{tab:NdepH}; the pressure was calculated without taking into account the long-range interaction.

\section{Conclusions \label{sec:concl}}
We performed molecular dynamics simulations of a nondegenerate hydrogen plasma using an improved Kelbg p/p that includes long–range interactions. The key difference from similar previous studies is an approximate treatment of the finite size of electrons with the same spin projection in their pair forces. This modification prevents the formation of nonphysical clusters containing a lot of bound electrons and protons. In this way, the Pauli exclusion principle is effectively enforced.

Verification of our technique on path integral Monte Carlo calculations by Filinov and Bonitz~\cite{Filinov:PhysRevE:2023} shows small deviations in energy and pressure for temperatures above 50~kK in the low degeneracy regime~$\chi \leq 0.01$. At lower temperatures, the differences grow and the energy becomes significantly underestimated. 

A similar behavior was reported earlier in Ref.~\onlinecite{Filinov:PhysRevE:2004} and explained by the formation of nonphysical clusters due to many-particle quantum exchange effects that were neglected. In our study, such clusters do not appear because the finite electron size is approximately accounted for in the interelectron forces. 

A further comparison of ionization degree and composition on the isotherm $T = 31250$~K indicates that the interparticle forces derived from the improved Kelbg p/p seem to produce excessive electron–proton attraction. This, in turn, leads to excessive formation of molecular compounds (while the Pauli principle holds). Our hypothesis is that, although the Kelbg p/p reproduces the exact density matrix accurately~\cite{Filinov:PhysRevE:2004}, its gradient can deviate notably from the exact one. A rigorous test of this hypothesis requires the calculation of the exact Coulomb density matrix and is beyond the scope of the present work.

Next, we performed MD simulations with a fixed degeneracy parameter of $\chi=0.01$. We computed radial distribution functions (RDFs) for a nondegenerate hydrogen plasma in the coupling range $0.1 \leq \Gamma \leq 3$, which corresponds to temperatures from 606~kK down to 670~K. As temperature decreases (or $\Gamma$ increases), a peak develops in the proton-proton RDF that reflects the formation of bound states, i.e., molecular compounds. At $\Gamma = 0.65$, the plasma becomes strongly diluted; as a result, there is a range of distances where the RDF is zero. For $\Gamma \geq 2$, a maximum followed by a minimum emerges at large distances, which indicates the onset of short-range order. We also found that the electron–electron RDF for the same spin projection decayed rapidly to zero at short distances. This behavior confirms that the Pauli principle is satisfied.

The analysis of ionization degree and composition in the nondegenerate regime shows an increase of the atomic fraction up to 50\% as $\Gamma$ reaches 0.5. Further increasing the coupling parameter results in the atomic fraction decreasing, while the molecular fraction grows. At $\Gamma = 1$, free electrons and protons vanish from the system. It is worth noting that the rapid and significant increase in molecular complexes may be an artifact of the improved Kelbg p/p. This finding should be validated through additional simulations that utilize the exact density matrix and its gradient.

We then obtained the thermodynamic limit for the energy and pressure of the nondegenerate hydrogen plasma. We found that the accounting for long-range interactions using the angular-averaged Ewald potential did not improve the $N$-convergence across the studied range of $\Gamma$. As $\Gamma$ increases and molecular species appear, the dependence on the number of particles becomes non-smooth for $N \leq 10^3$. Because of this fact, a fit of energy $N$-dependence over all points is possible only for $\Gamma \leq 0.5$. A likely solution is to use simulations with a variable number of particles that is beyond the scope of this work.

Finally, we provided the tabulated energy and pressure in the thermodynamic limit for the nondegenerate hydrogen plasma in the range $0.1 \leq \Gamma \leq 2.5$ at $\chi = 10^{-2}$.

\begin{acknowledgments}
The work was supported by the Foundation for the Advancement of Theoretical Physics and Mathematics ``BASIS'' (Grant No. 23-1-5-119-1).
We thank Alexander Onegin for testing and suggestions to speed up calculations in LAMMPS.
The authors acknowledge the JIHT RAS Supercomputer Centre, and the Shared Resource Centre ``Far Eastern Computing Resource'' IACP FEB RAS for providing computing time.
\end{acknowledgments}

\section*{Data Availability Statement} The data that support the findings of this study are available within the article.

\appendix
\section{Temperature derivatives of pseudopotentials \label{app:1}}
In this section, we present formulas for the temperature derivatives of the contributions to the internal energy~\eqref{eq:fullPotEnergyQuasiClMD}. First, we consider the derivative of the contribution that accounts for long-range interaction.
The derivative with respect to temperature at $r_{ij} \leq r_m$ has the form (we use the notations $x_{ij} = r_{ij}/\lambda_{ij}$ and $x_m = r_m/\lambda_{ij}$):
\begin{multline}
	\label{eq:rdgdgfrf}
	\beta \Bigl(\frac{\partial \Phi_1({r}_{ij};\beta)}{\partial \beta}\Bigr)_V
	= \tfrac{1}{128 r_{ij} x_m^3}
	\Big[ 
	\sqrt{\pi}\Bigl(4\Bigl(x_{ij}^4 -3x_{ij}^2(2x_m^2+1)\\+8x_{ij}x_m^3 -3x_m^4\Bigr) 
	+12x_m^2 -9\Bigr)\,\erf(x_{ij}-x_m)
	\\+\sqrt{\pi}\Bigl(4\Bigl(x_{ij}^4 -3x_{ij}^2(2x_m^2+1)-8x_{ij}x_m^3 -3x_m^4\Bigr) 
	\\+12x_m^2 -9\Bigr)\,\erf(x_{ij}+x_m)
	+2\,e^{-(x_{ij}+x_m)^2}\Bigl(2x_{ij}^3 -2x_{ij}^2 x_m 
	\\- x_{ij}(10x_m^2+7)-6x_m^3 +9x_m\Bigr) 
	+2\,e^{-(x_{ij}-x_m)^2}\Bigl(2x_{ij}^3 +2x_{ij}^2 x_m 
	\\- x_{ij}(10x_m^2+7)+6x_m^3 -9x_m\Bigr) 
	+64\,x_{ij}\,(\sqrt{\pi}\,x_m^3+1) 
	\Big].
\end{multline}

Let us find the temperature derivative of the improved Kelbg p/p. Since $\Phi^\text{I}_0({r}_{ij}, \beta)=\Phi^\text{I}_0({r}_{ij}, \lambda_{ij}(\beta), \gamma_{ij}(\beta))$, then
\begin{multline} 
	\label{eq:frbdgrsvgdbg} 
	\beta \frac{\partial \Phi^\text{I}_0({r}_{ij}, \beta)}{\partial \beta} 
	= 
	\frac{\partial \Phi^\text{I}_0({r}_{ij}, \lambda_{ij}(\beta), \gamma_{ij}(\beta))}{\partial \lambda_{ij}(\beta)} 
	\frac{\lambda_{ij}}{2} 
	\\
	+ \frac{\partial \Phi^\text{I}_0({r}_{ij}, \lambda_{ij}(\beta), \gamma_{ij}(\beta))}{\partial \gamma_{ij}(\beta)} 
	\beta 
	\frac{\partial \gamma_{ij}(\beta)}{\partial \beta},
\end{multline}
where all the derivatives are at fixed volume $V$.
Let's write out all the derivatives in Eq.~\eqref{eq:frbdgrsvgdbg}:
\begin{multline} 
	\frac{\partial \Phi^\text{I}_0({r}_{ij}, \lambda_{ij}(\beta), \gamma_{ij}(\beta))}{\partial \lambda_{ij}(\beta)} 
	= 
	\frac{2 r_{ij}}{ 
		\lambda_{ij}^3}  ( 
	e^{ -\frac{ \gamma_{ij}^2 r_{ij}^2 }{ \lambda_{ij}^2 } } 
	- e^{-\frac{ r_{ij}^2 }{ \lambda_{ij}^2 }  }	) \\
	{} - \frac{ \sqrt{\pi} \, \lambda_{ij} \, \erfc\left( \frac{ \gamma_{ij} r_{ij} }{ \lambda_{ij} } \right) }{ \gamma_{ij} \lambda_{ij}^3} 
	,
\end{multline}
\begin{multline} 
	\frac{\partial \Phi^\text{I}_0({r}_{ij}, \lambda_{ij}(\beta), \gamma_{ij}(\beta))}{\partial \gamma_{ij}(\beta)} 
	= 
	- 
	\frac{\sqrt{\pi} \, \lambda_{ij}}{ 
		\gamma_{ij}^2 \lambda_{ij}^2 
	} \, \erfc\left( \frac{ \gamma_{ij} r_{ij} }{ \lambda_{ij} } \right) \\
	{} - \frac{2 \gamma_{ij} r_{ij}}{ 
		\gamma_{ij}^2 \lambda_{ij}^2 
	} \, e^{ -\frac{ \gamma_{ij}^2 r_{ij}^2 }{ \lambda_{ij}^2 }}
	.
\end{multline}
In the case of interaction between an electron and a proton (see equation (22) in Ref.~\onlinecite{Filinov:PhysRevE:2004}):
\begin{equation}
	\beta
	\frac{\partial \gamma_{ep}(\beta)}{\partial \beta}
	=
	- \frac{x_\beta}{2} \times
	\frac{1 + 2x_\beta + (-1 + a_{\text{ep}}) x_\beta^2}
	{(1 + a_{\text{ep}} x_\beta + x_\beta^2)^2}.
\end{equation}
In the case of interaction between electrons (see equations (23) and (24) in Ref.~\onlinecite{Filinov:PhysRevE:2004}):
\begin{multline}
	\beta
	\frac{\partial \gamma_{ee}(\beta)}{\partial \beta} 
	=
	\frac{1}{6 (1 + x_{\beta}^2)^2}
	\Bigg[
	3 x_{\beta} 
	\Bigl(
	2(-1 + \gamma_{ee,0}) x_{\beta} 
	+\\ a_{ee} (-1 + x_{\beta}^2)
	\Bigr)
	+
	\frac{
		2 (1 + x_{\beta}^2) \tilde{x}_{\beta}^3
		\left(
		4 + 3\tilde{x}_{\beta}^2
		- 3 \ln\left( \frac{8 \tilde{x}_{\beta}^4}{\sqrt{\pi}} \right)
		\right)
	}{
		\sqrt{\pi}
		\left(
		\ln\left( \frac{8 \tilde{x}_{\beta}^4}{\sqrt{\pi}} \right)
		- 3 \tilde{x}_{\beta}^2
		\right)^2
	}
	\Bigg] .
\end{multline}


%

\end{document}